\documentclass[iop,apj,numberedappendix,twocolumn, tighten]{aastex62}

\usepackage{arydshln}
\usepackage{natbib}
\usepackage{graphicx}
\usepackage{amsmath,amsfonts}
\usepackage{multirow}
\usepackage{xspace}
\usepackage[normalem]{ulem}
\usepackage{booktabs}
\usepackage{colortbl}
\usepackage{hyperref}
\usepackage[nameinlink]{cleveref}
\usepackage[encapsulated]{CJK}
\usepackage{ucs}
\usepackage{float}
\usepackage[utf8]{inputenc}

\usepackage{eht}
\usepackage{newtxtext}
\usepackage{tablefootnote} 
\usepackage{mathrsfs}
\usepackage{lineno}
\usepackage{bbm}

\definecolor{fed_blue}{HTML}{07004D}
\definecolor{steel_blue}{HTML}{2D82B7}
\definecolor{steel_blue_dark}{HTML}{1C71A6}
\definecolor{aqua_marine}{HTML}{42E2B8}
\definecolor{dutch_white}{HTML}{F3DFBF}
\definecolor{light_coral}{HTML}{EB8A90}
\definecolor{light_coral_dark}{HTML}{BA5A60}

\hypersetup{
    colorlinks = true,
    linkcolor = purple,
    urlcolor  = steel_blue_dark,
    citecolor = steel_blue_dark,
    anchorcolor = black
}

\graphicspath{{./figures}}

\newcommand{\uv}{$(u,v)$\xspace}

\def\mring{$m$-ring\xspace}
\def\lsim{\mathrel{\raise.3ex\hbox{$<$\kern-.75em\lower1ex\hbox{$\sim$}}}}
\def\gsim{\mathrel{\raise.3ex\hbox{$>$\kern-.75em\lower1ex\hbox{$\sim$}}}}
\def\gtwid{\mathrel{\raise.3ex\hbox{$>$\kern-.75em\lower1ex\hbox{$\sim$}}}}
\def\proptwid{\mathrel{\raise.3ex\hbox{$\propto$\kern-.75em\lower1ex\hbox{$\sim$}}}}

\defcitealias{TMS}{TMS}
\setcitestyle{notesep={; }}

\begin{document}

\title{\textbf{The First 4 Years of SN 1993J Revisited: Geometric $m$-ring Modeling of the Radio Shell with Closure Quantities Only}} 
\shorttitle{SN 1993J Geometric Analysis}

\author[0000-0003-4914-5625]{Joseph R. Farah}
\affiliation{Las Cumbres Observatory, 6740 Cortona Drive, Suite 102, Goleta, 
CA 93117-5575, USA}
\affiliation{Department of Physics, University of California, Santa Barbara, 
CA 93106-9530, USA}

\author[0000-0002-7174-8273]{Logan J. Prust}
\affiliation{Kavli Institute for Theoretical Physics, University of California, Santa Barbara, CA, USA}

\author[0000-0003-0794-5982]{Giacomo Terreran}
\affiliation{Las Cumbres Observatory, 6740 Cortona Drive, Suite 102, Goleta, 
CA 93117-5575, USA}
\author[0000-0003-4253-656X]{D. Andrew Howell}
\affiliation{Las Cumbres Observatory, 6740 Cortona Drive, Suite 102, Goleta, 
CA 93117-5575, USA}
\affiliation{Department of Physics, University of California, Santa Barbara, 
CA 93106-9530, USA}

\author[0000-0002-0592-4152]{Michael Bietenholz}
\affiliation{Department of Physics and Astronomy, York University, 4700 Keele St., Toronto, M3J 1P3, Ontario, Canada}
\author{Norbert Bartel}
\affiliation{Department of Physics and Astronomy, York University, 4700 Keele St., Toronto, M3J 1P3, Ontario, Canada}
\author[0000-0001-5807-7893]{Curtis McCully}
\affiliation{Las Cumbres Observatory, 6740 Cortona Drive, Suite 102, Goleta, 
CA 93117-5575, USA}
\author[0000-0002-4120-3029]{Michael D. Johnson}
\affiliation{Black Hole Initiative at Harvard University, 20 Garden Street, Cambridge, MA 02138, USA}
\affiliation{Center for Astrophysics | Harvard \& Smithsonian, 60 Garden Street, Cambridge, MA 02138, USA}


\shortauthors{Farah et al.}

\correspondingauthor{$^\dag$Joseph R. Farah}
\email{josephfarah@ucsb.edu}

\begin{abstract}
SN 1993J is the best-studied radio supernova, with observations using very-long-baseline interferometry (VLBI) spanning from within weeks of explosion through nearly three decades of ejecta evolution. Imaging and modeling techniques have revealed an expanding shell-like ejecta structure, with a width well-constrained after $\sim1000$ days. In this work, we present a re-analysis of a portion of the first $\sim1700$ days of SN 1993J evolution, using a new VLBI analysis technique with somewhat intrinsically higher angular resolution and compare our results with those from previous conventional techniques. We adopt the \mring model, with potentially somewhat higher angular resolution but only as a phenomenological alternative to the physically-motivated spherical shell model used in past analyses and shape the profile to approximately match that of the shell. The \mring model allows us to fit the ejecta diameter, width, central opacity, and azimuthal brightness modulation simultaneously. Additionally, we use closure quantities only, providing calibration insensitive constraints on ejecta geometry and largely independent comparisons with previous results from the same data sets. Using this approach we find the ejecta expanding with a power-law exponent $\omega=0.80\pm0.01$ averaged for the time from 175 to 1693 days, consistent with shell-fitting analyses.
For the first time, we report estimates of the width of the ejecta at $t \lesssim 1000$ days—as early as 264 days  post-explosion—finding a mean fractional width with standard deviation of  $0.24 \pm 0.04$ (of the radius) with no significant evolution. Further, we present a fit of the azimuthal brightness modulation over time with the maximum of an apparent horseshoe pattern rotating from east to south-southwest from 175 to 1000 d and then evolving to a more complex behavior, quantifying previous results from images only. The constraints on the angular brightness modulation and width over time and $\sim700$ days earlier than existing analyses provide an avenue to test more advanced simulations of the explosion and interaction mechanisms.

\end{abstract}


\section{Introduction}
\label{sec:introduction}

Supernova (SN) 1993J was discovered in March 1993 in the spiral galaxy M81 \citep{Disc_93J}. As the prototypical Type IIb supernova \citep{Filippenko1993}, it was intensely studied across a variety of wavelengths, substantially facilitated by its proximity \citep[3.6 Mpc;][]{M81_Distance} and peak brightness \citep{Weiler2007}. In particular, the radio-bright nature of the explosion led to a decade-long follow-up radio campaign using VLBI \citep[see e.g.,][for a review]{Marcaide2009}, which resolved an expanding asymmetric quasi-spherical shell \citep{Bartel1994,Marcaide1995,Bietenholz2001}. The direct observation of the radio shell expansion provides a unique opportunity to probe high-energy supernova physics at all initial phases of the explosion. 

Previous works have characterized the behavior of the expanding shell using a variety of approaches. Both modeling and imaging approaches have constrained the apparent size and width of the shell-like structure, and observed a decelerating power-law expansion \citep{Marcaide1997,Bartel2000,Bartel2002} with a fractional shell width of $\approx30\%$ \citep{Marcaide1995SHELL,Bartel2000,Bietenholz2005,Marcaide2009}. In particular, the shell width at early epochs ($\lesssim1000$ days) has not been well-characterized. Images at these early epochs are beam-convolved and cannot be used to measure a shell width, while width-enabled model-fits either fixed the width prior to day 900 \citep{Bartel2002} or did not fit width and size independently \citep{Marcaide2009}. The reported shell width is consistent with previous simulations and theoretical predictions \citep[see e.g.,][etc.]{Chevalier1978,Kundu2019}. However, existing simulations of the SN 1993J explosion have been too simplistic (i.e., not including the impact of Rayleigh-Taylor instabilities or the complex interaction with the circumstellar medium, abbr. CSM) to present a robust prediction for the width of the emission region.

Multiple imaging investigations have attempted to characterize the degree and evolution of asymmetry present in the azimuthal brightness along the ridge of the shell, which report a spherically symmetric object with an azimuthal brightness modulation \citep{Bartel1994,Marcaide1995,Bietenholz2001,Bietenholz2003}. In particular, \cite{Bietenholz2003} found a horseshoe-like brightness pattern in the images with the maximum rotating from east to south south-west over the first 1000 d before dissolving into a more complex structure. Concerns have been raised that the reported asymmetry and subsequent evolution may be artifacts caused by uneven and evolving \uv coverage \citep{Heywood2009}. Existing models \citep[e.g., a spherical shell of uniform emissivity;][]{Bartel2000} have been symmetric; a major limitation of such models are that they cannot characterize any asymmetry in intensity or shape present in the data. Recently, the $m$-ring model of \cite{Johnson2019} has been developed which varies the size, width, and brightness asymmetry of a ring. The $m$-ring model (an infinitesimally thin ring blurred with a Gaussian kernel) looks phenomenologically similar to the SN 1993J images, but is not as physically motivated as the spherical shell model of supernovae ejecta. However, it has the benefit of incorporating azimuthal brightness variations of arbitrary complexity while retaining an analytic Fourier transform. Such functionality allows for a more extensive investigation of the impact of arbitrary azimuthal brightness modulations on estimates of size and width, as well as novel estimates (with uncertainties) of the ejecta brightness asymmetry independent of imaging.

A major challenge in the study of SN 1993J has been assessing the impact of calibration choices and other systematics on imaging and model inference. Previous works have investigated the impacts of model assumptions \citep{Bartel2002} and data processing \citep{Marcaide2009}. The effect on inferred parameters and reconstructed images is significant, resulting in conflicting measurements from different analyses. Previous analyses have attempted to address this issue, reanalyzing the full dataset and attempting more careful self-calibration \citep{Vidal2024}. Recently, \cite{Chael2018} and others explored imaging with so-called ``closure'' quantities, derived quantities which by construction are invariant to phase errors, complex gains, and several other systematic artifacts. Directly modeling the SN 1993J evolution using these quantities may sidestep particular calibration and systematic error concerns entirely, although the amount of independent information would decrease and the total flux density information would be lost.

In this paper, we re-analyze selected data sets of the first four years of the evolution of the SN 1993J radio shell, using a new geometric model capable of measuring asymmetry and basing our inference solely on calibration-insensitive closure quantities. In \autoref{sec:data}, we summarize the observation and subsequent reduction of the data taken on the SN 1993J explosion. In \autoref{sec:mring_modeling_and_fitting_framework}, we present our modeling scheme and fitting approach, and detail how we assess fit performance and quality. In \autoref{sec:application_to_sn_1993j}, we apply our modeling framework to the radial evolution of SN 1993J, and present the first-ever quantitative measurements of asymmetry in the azimuthal brightness variation along the ridge of the ring, as well as the first width measurements $\lesssim900$ days post-explosion. Finally in \autoref{sec:discussion} and \autoref{sec:conclusions}, we discuss our results and methods, and summarize our conclusions. 
\section{Overview of SN 1993J data products}
\label{sec:data}

\subsection{Observations and data reduction}
\label{sub:observations_and_data_reduction}

SN 1993J was observed 22 times at 8.4 GHz over the range of interest for our analysis, between explosion in 1993 and 1997. The observations were conducted at varying times of year, multiple times each year. The earliest observation is estimated to correspond to 50 days post-explosion. The latest observation is estimated to correspond to $\sim$1700 days post-explosion. For a full breakdown of the array and contributing interferometers and stations, see \cite{Bietenholz2001,Bartel2002}. We list stations we reference in this paper and their associated acronyms in \autoref{tab:acronym}.

Each observation was conducted at $\approx$8.4 GHz, corresponding to a wavelength of $\approx$3.6 cm. Between 100 and 500 scans were performed over each observation. The number of stations available to observe ranged from 12 to 18 stations, which produced $n(n-1)/2$ baselines ranging from 66 to 153 baselines, $\approx \binom{n}{3}$ closure triangles ranging from 220 to 816 triangles, and $\approx \binom{n}{4}$ closure quadrangles ranging from 495 to 3060 quadrangles.

The observations were phase-referenced using the galactic core of M81 as a calibrator, alternating between SN 1993J and M81 every few minutes. The proximity of SN 1993J to the M81 core makes M81 an excellent calibrator, enabling the production of high-fidelity images at both early epochs (when the small size of the source presents a challenge) and late epochs (when the low flux density of the source presents a challenge). 

\begin{table}[ht]
\centering
\caption{A list of VLBI network sites referenced in this paper and their corresponding two-letter acronyms.}
\begin{tabular}{ll}
\hline
\textbf{Acronym} & \textbf{Station Name} \\
\hline
OV  & Owens Valley Radio Observatory, California \\
NT  & Noto Radio Observatory, Italy \\
RB  & Robledo de Chavela, Spain \\
SC  & Saint Croix, U.S. Virgin Islands \\
HN  & Hancock, New Hampshire \\
Y   & Very Large Array (VLA), New Mexico \\
LA  & Los Alamos, New Mexico \\
NL  & North Liberty, Iowa \\
PT  & Pie Town, New Mexico \\
BR  & Brewster, Washington \\
FD  & Fort Davis, Texas \\
MK  & Mauna Kea, Hawaii \\
KP  & Kitt Peak, Arizona \\
\hline
\end{tabular}
\label{tab:acronym}
\end{table}

\subsection{Construction of closure quantities}
\label{sub:construction_of_closure_quantities}

Here, we review the construction of various VLBI data products. We primarily cite \cite{Thompson2001} for general conventions and descriptions, and \cite{Chael2018} and \cite{LindyClosure} for expressions and descriptions related to the construction of closure quantities. 

Interferometers measure a quantity known as a visibility $V_{ij}$, which is probed between two stations $i$ and $j$, forming the baseline $b_{ij}$. This visibility measures the complex Fourier component of the source image by the van Cittert-Zernike theorem
\begin{align}
    V_{ij} = \mathcal{F}_{uv} I(x, y) = \int dx \ dy \ I(x, y) e^{2\pi \mathbbm{i} (ux + vy)}.\,
    \label{eq:vancittertzernike}
\end{align}
Here $I(x, y)$ is the intensity distribution of the source image and $\mathcal{F}_{uv}$ is the Fourier transform which takes spatial components $(x, y)$ to frequency space $(u, v)$. The coordinates $(u, v)$ represent the components of $b_{ij}$ projected along the line-of-sight from the array to the source. By transforming models from image space into Fourier space, we can compare predicted (model) visibilities to observed (data) visibilities and identify models which best reproduce the observations. 

The observed visibility is corrupted from \autoref{eq:vancittertzernike} by several sources of error:
\begin{itemize}
    \item \textit{Thermal noise}: thermal noise $\epsilon_{ij}$ is added to the true visibility from a combination of Earth's atmosphere, non-zero telescope temperature, and an astrophysical background. The thermal noise is Gaussian, and varies both with time and the specific baseline being considered.
    \item \textit{Phase error}: the argument of the complex visibility is modified due to corrupting effects from the atmosphere and instrumentation/timing errors. Site-specific phase error $\phi_i$ is incorporated as a phase shift in the true visibility.
    \item \textit{Complex gains}: the visibility is further modified by a complex gain $G_i$ which is incorporated multiplicatively. The complex gain is time-dependent and varies on a per-site basis. 
\end{itemize}
Following \cite{Chael2018}, we ignore more complex corrupting effects, such as per-station polarization mixing. Incorporating these effects into the visibility gives,
\begin{align}
    V_{ij}' = G_i G_j e^{\mathbbm{i}(\phi_i - \phi_j)}(V_{ij} + \epsilon_{ij}).
\end{align}
These errors on the visibility complicate model inference, as they corrupt the true visibility. Previous analyses of SN 1993J and other VLBI targets have primarily focused on self-calibration, which uses redundancy in the data to iteratively correct phase and amplitude errors. Recently, \cite{Chael2018} demonstrated that a set of derived quantities called ``closure quantities'' are not affected by many of the most significant sources of error and can be employed to create robust and gain-insensitive images. In this work, we seek to exploit two closure quantities (the closure phase and the log closure amplitude) to model the SN 1993J ejecta, independent of complex gain magnitude or calibration choices. 

\subsubsection{Closure phases}
\label{ssub:closure_phases}

The closure phase quantity is constructed by adding the phases of 3 baselines (forming a triangle) in closure. The closure phase is the argument of the complex bispectrum $V_B = V_{12}V_{23}V_{31}$, 
\begin{align}
    \phi_{123} = \arg{V_B} = \arg V_{12} + \arg V_{23} - \arg V_{13}.
\end{align}
This construction has the benefit of only being sensitive to thermal noise and is robust to complex gains and phase errors. As a result, the closure phases computed for an observation will be identical to the closure phase computed for the true source image (modulo contributions from thermal noise). Since the closure phases are sensitive only to thermal noise, we use the method of \cite{Wielgus2019EHT} to estimate the present thermal noise from closure phases and appropriately rescale the measured uncertainties. We note that this procedure will underestimate the noise present in the data for several reasons: (i) the expressions in \cite{Wielgus2019EHT} are valid at high $S/N$; however, much of the SN 1993J data (particularly at later days) is at low $S/N$; (ii) there may be additional noise sources besides thermal noise present; and (iii) we assume Gaussian uncertainties when at least the low $S/N$ data likely follow a Ricean distribution. All fits are performed using these rescaled uncertainties. 

\subsubsection{Log closure amplitudes}
\label{ssub:log_closure_amplitudes}

The closure amplitude is formed by taking the ratio of the product of visibilities on one pair of baselines to the product on a second pair, and then taking the absolute value to remove phase dependence:
\begin{align}
    |V_{\rm LCA}| = \left|\frac{V_{12}V_{34}}{V_{13}V_{24}}\right|.
\end{align}
All possible cyclic permutations then form the set of closure amplitudes for a single quadrangle of four antennas. The effect of this construction is to cancel out the complex time-dependent gain terms $G_i$, resulting in a robust data product. 
At low signal-to-noise (S/N), the closure amplitude distribution becomes substantially non-Gaussian due to the $1/|V_{13}V_{24}|$ term, complicating estimates of the likelihood. The log closure amplitude (constructed by taking the natural log of the closure amplitude) addresses the non-Gaussian tail of the reciprocal visibility amplitude and has the added benefit of avoiding the preferential weighting of the denominator baselines. \cite{Chael2018} and \citep{LindyClosure} found that the log closure amplitude was a more robust observable for imaging and compatible with a Gaussian likelihood distribution, even at lower S/N. 

\subsubsection{Closure quantity special considerations}

Use of closure quantities in imaging and modeling has several nuances, to which we describe our approach here:

\begin{enumerate}
    \item \textit{Closure quantities formed using zero baselines}: in some interferometers, stations may be so close together geographically that the baseline between them does not meaningfully probe source image structure on scales of interest (a ``zero baseline''). Triangles formed with a zero baseline have zero closure phase, which do not add new information but can be used to improve measurements via averaging \citep{Fish2016}. Geographically co-located sites also result in trivial closure amplitudes, but baselines to co-located sites can add new information \citep{Johnson2015}. To maximize available information and robustness, we follow \cite{Chael2018} and include closure quantities constructed using zero baselines. 
    \item \textit{Diagonalized closure quantities}: recent work \citep{LindyClosure} has presented diagonalized closure phases and log closure amplitudes, which transform closure quantities into an orthogonal basis. This transformation has the effect of negating possible correlations between measurements. All fits performed in this study use diagonalized closure quantities.
    \item \textit{Debiasing of log closure amplitudes}: in the low $S/N$ regime, the presence of noise on positive definite amplitude measurements (from which the closure amplitude is constructed) results in a Ricean bias which biases the amplitudes to larger values. Amplitudes can be debiased using the first order correction presented in \cite{Thompson2001}:
    \begin{align}
        |V_{ij}|_{\textrm{debiased}} = \left[ |V_{ij}|^{2}_{\textrm{measured}} - \sigma_{ij}^2\right]^{1/2}, 
    \end{align} 
    where $\sigma_{ij}$ is the standard deviation of the thermal noise on $b_{ij}$. Following \cite{Chael2018}, we debias all closure amplitudes prior to fitting. However, even with debiasing, if the data have a high fraction of data points with $S/N < 1$, model fits will still be Ricean biased. We investigate this and find that on $t \lesssim 1356$ d, the percentage of points with $S/N < 1$ is $\lesssim10\%$, which our debiasing procedure can sufficiently accommodate. For $t=1356, 1532$ and $1693$ d, the percentage of points with $S/N < 1$ is $\sim10\%$, leading to a possible minor bias in the model fits at these epochs. We choose to cautiously include these days in our analysis, but fit results from these days should be interpreted with this possible source of bias in mind.
    \item \textit{Total intensity constraint}: it is understood that the closure phase and closure amplitude cannot recover information about the total intensity of a source \citep{Chael2018}. Therefore, we perform our model fits in this paper with a fixed total intensity. The total intensity is chosen to be the average of the intensity on the shortest baselines ($\rho \equiv \sqrt{u^2 + v^2} < 0.02$ G$\lambda$). 
\end{enumerate}

\section{$m$-ring modeling and fitting framework}
\label{sec:mring_modeling_and_fitting_framework}

\subsection{Construction of the $m$-ring model}
\label{sub:construction_of_the_m_ring_model}

Here, we motivate and construct a model which we will use to fit the SN 1993J data quantities. Previous imaging-based investigations \citep[see e.g.,][]{Marcaide2009,Bietenholz2005} of the SN 1993J ejecta have identified the presence of a spherical shell expanding with time. The spherical shell morphology, when projected into the two-dimensional observer plane, can be well-approximated by a thick ring model with uniform brightness in the center.  We propose a generic, single-component model based on the \mring model presented in \cite{Johnson2019}---specifically equations (18) and (19), which we reproduce here. The \mring considers a ring of angular diameter $d$, azimuthally-varying brightness, and infinitesimal width. The brightness distribution around the ring is formulated as a sum of polar Fourier modes, giving the following polar intensity function,
\begin{equation}
    I\left(\rho, \varphi_{\rho}\right)=\frac{1}{\pi d} \delta\left(\rho-\frac{d}{2}\right) \sum_{m=-\infty}^{\infty} \beta_{m} e^{i m \varphi_{\rho}}.
\end{equation}
Here, $\beta_m$ is the coefficient corresponding to the $m$-th Fourier mode, with $\beta_0 > 0$ representing the total flux density of the ring. The resulting visibility function is
\begin{equation}
    V\left(u, \varphi_{u}\right)=\sum_{m=-\infty}^{\infty} \beta_{m} J_{m}(\pi u d) e^{i m\left(\varphi_{u}-\pi / 2\right)}.
    \label{sec:modeling:eq:visibities}
\end{equation}
Here, the function $J_m$ refers to the $m$-th Bessel function of the first kind, which can be approximated by
\begin{equation}
    J_{m}(\pi u d) \approx \frac{1}{\pi} \sqrt{\frac{2}{u d}} \cos \left[\pi\left(u d-\frac{2 m+1}{4}\right)\right].
\end{equation}
Opacity effects (causing the brightness in the center of the shell projection to be non-zero) are reproduced in the \mring model by the addition of a ``filling floor'', a disk of constant brightness and total flux density $F_d$ with the same radius as the \mring. The brightness of the disk is controlled in the model input by the ``fractional filling floor'' parameter $f$, which represents the fraction of the total flux density $F_d/(F_d+\beta_0)$ the filling floor contributes. In general, the spherical shell model is well-approximated by $f\in[0.3, 0.45]$. We note that the log closure amplitude and closure phase do not constrain the total intensity of the source, and so we do not fit this parameter and instead (for visualization or calculation purposes) fix it to the average amplitude of the shortest baselines ($\rho < 0.02$ G$\lambda$).  
The model is then convolved with a Gaussian kernel of full-width half max (FWHM) $\alpha$ to produce a thickness. Finally, since closure quantities are insensitive to offsets of the source image from the origin, we do not incorporate any translation parameters into our model. Examples of the model parameters are shown in \autoref{tab:mring_params}.

\begin{table*}[!ht]
\centering
\begin{tabular}{p{0.12\linewidth} p{0.60\linewidth} p{0.22\linewidth}}
\toprule
\textbf{Parameter} & \textbf{Explanation} & \textbf{Example values} \\
\midrule
$I_0$ & Total flux of the model, split between the ring and the central floor (not constrained by closure quantities) & 1 Jy, 0.1 Jy \\
$d$ & Diameter of both the \mring{} and the disk & 1 mas, 500~\textmu{}as \\
$\alpha$ & Blurring kernel that produces a ring width of $\alpha$ & 500~\textmu{}as, 0.1 mas \\
$f$ & Fraction of the total flux from the central disk & 0.1, 0 \\
$\beta_m$ & Complex value describing azimuthal flux distribution: $|\beta_1|$ sets asymmetry strength; $\arg\beta_1$ sets direction & 0, $0.2e^{i\pi/3}$, $0.2 + 0.17i$ \\
\bottomrule
\end{tabular}
\caption{A list of \mring{} model parameters, their definitions, and sample values.}
\label{tab:mring_params}
\end{table*}

\begin{figure}
    \centering
    \includegraphics[scale=0.6]{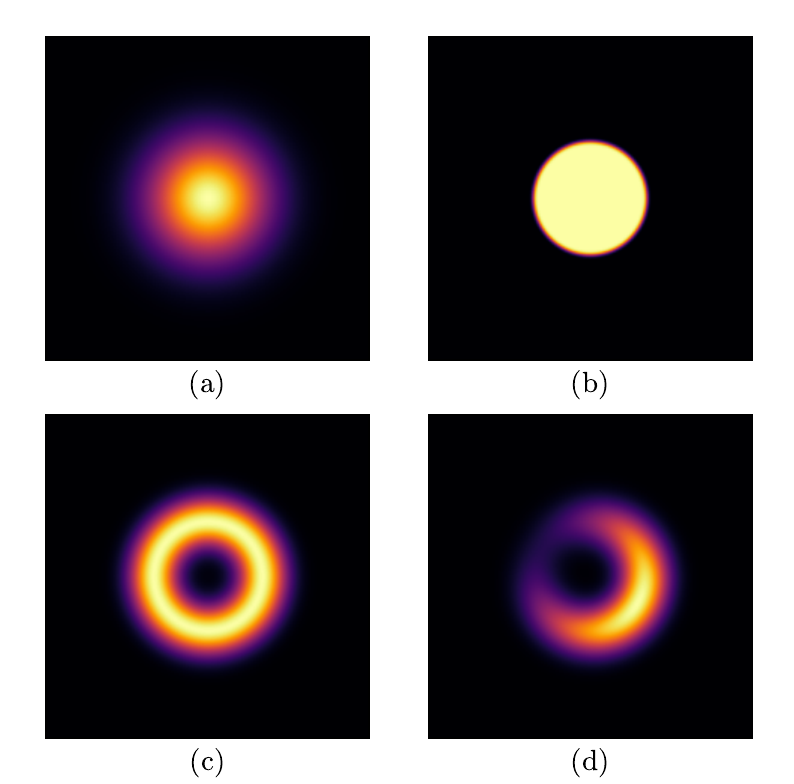}
    \caption{A demonstration of the different morphologies within the \mring model. (a) A Gaussian morphology, produced by an \mring model with a small diameter parameter $d$ and large blurring kernel parameter $\alpha \gg d$. (b) A disc morphology, produced by an \mring model with a small blurring kernel parameter $\alpha \ll d$ and a 100\% fractional floor. (c) A ring morphology, produced by a 0\% fractional floor and a blurring kernel parameter $\alpha{\sim} d$ comparable to the diameter. (d) A crescent morphology, produced by giving a non-zero significance to the cosine-expansion azimuthal ring distribution. Note that all these models are circularly symmetric, in that they are constructed by varying a brightness distribution on a circle; thus, even in the case of the crescent (d), though the brightness varies around the ring, the width of the Gaussian radial profile does not.}
    \label{fig:mring_demo}
\end{figure}

This model is convenient for several reasons. First, the visibility function is analytic, which is important for improving performance in rigorous but slow fitting algorithms such as Markov-Chain Monte Carlo. Additionally, it is a single-component model which can transition seamlessly from Gaussians to disks to rings to crescents, which covers a large range of circularly-symmetric morphologies we expect to be represented in the early SN 1993J data. Examples of the different morphologies reproducible in this model are shown in \autoref{fig:mring_demo}. 

\subsection{Epoch selection criteria}
\label{sub:epoch_selection_criteria}

To maximize the reliability of our fits, we exclusively fit to a selection of the dataset which favors our methods and models. First, while the $m=1$ \mring model can reproduce a variety of model morphologies (namely, discs, rings, Gaussians, and crescents), it may perform poorly if the ejecta asymmetry is more complex than a simple one-mode angular brightness profile. Imaging \citep{Bietenholz2005,Marcaide2009} indicates that for $t \lesssim 1000$ d, the ejecta are well-reproduced by the one-mode \mring model, as demonstrated in \autoref{fig:mring_demo}. However, for $t\gtrsim 1000$ d, the angular brightness profile becomes substantially more complex as the ejecta breaks up under the ram pressure of the surrounding gas. To mitigate this limitation of the \mring model, we use the $|\beta_1|$ parameter as a diagnostic for data-model compatibility, as $|\beta_1| > 0$ indicates the ejecta retain some degree of $m=1$ mode asymmetry. By contrast, $|\beta_1| \sim 0$ indicates that the ejecta are approximately symmetric in brightness at the $m=1$ mode level, requiring $m>1$ modes to properly capture the azimuthal variation. Empirically, our modeling scheme reports $|\beta_1|$ first becomes consistent with zero up to $1\sigma$ at $t = 1693$ d (see \autoref{tab:results}); thus we restrict our dataset to $t \leq 1693$ d. Indeed, we find that for $t=1356,1532$, and $1693$ d---where $|\beta_1|$ is consistent with zero up to $2\sigma$ or better---recovery of $\arg\beta_1$ is highly uncertain (uncertainties of $\pm 90^\circ, \pm 159^\circ, \pm 96^\circ$ respectively, $1\textrm{-}2$ orders of magnitude higher than the rest of the dataset), indicating the notion of the brightness asymmetry that the $m=1$ mode probes has become less meaningful. In the future, we will explore increasing the complexity of the \mring model; however, for the purposes of this analysis (which aims to characterize the mean asymmetry direction of the ejecta), we fit to $t\lesssim 1500$ d in order to maximize compatibility with the \mring model. 

\begin{figure}
    \centering
    \hspace*{-1cm}\includegraphics[scale=0.7]{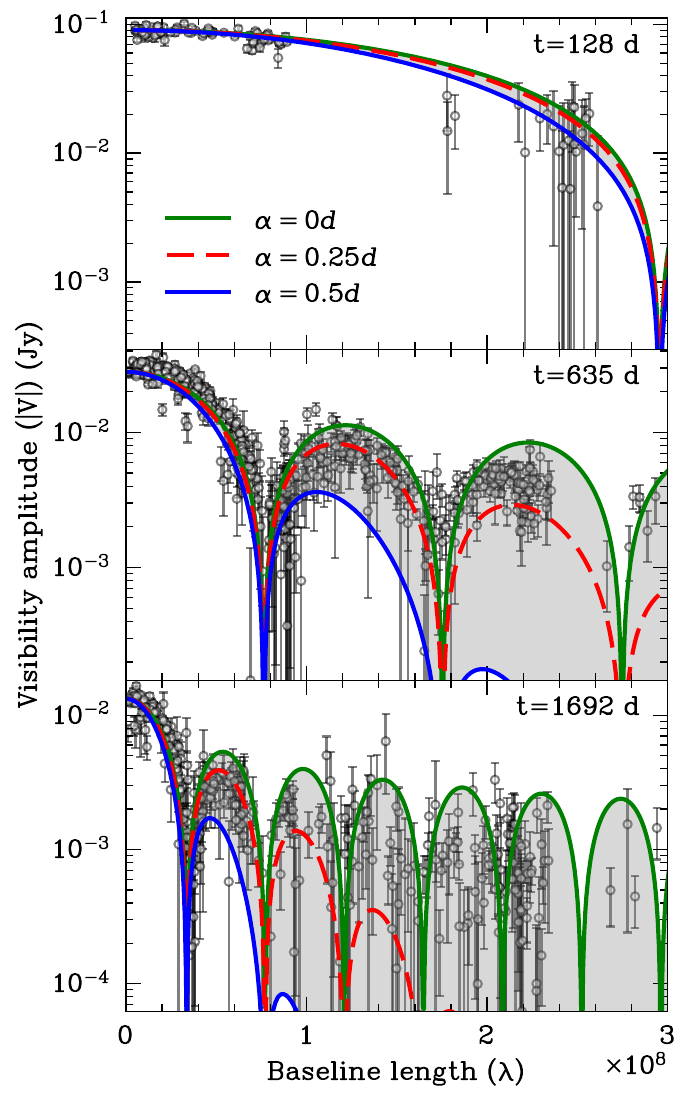}
    \caption{Visibility amplitudes for several days in the SN 1993J dataset, with the corresponding $m$-ring Fourier transform. The $m$-ring width is varied between infinitesimally thin (green solid line) to unphysically thick (blue solid line), with an intermediate value of $\alpha=0.25d$ shown for demonstration purposes (red dashed line). We can only meaningfully constrain the width of the ejecta structure as long as the second peak (first peak is at $\rho=0$) is clearly visible; earlier than this, the ejecta are effectively degenerate with a bivariate Gaussian intensity distribution. Thus, we only fit a width to $t \gtrsim 264$ d, the earliest day when both the first null and the second peak are visible. As a result, we are able to constrain the width $\gtrsim 700$ days earlier than previous works.}
    \label{fig:width_amp_demo}
\end{figure}

Second, our model incorporates a width parameter $\alpha$, and fits both $\alpha$ and $d$ simultaneously. Previous analyses indicated difficulty fitting width and diameter simultaneously \citep{Marcaide2009}; we do not find such difficulty in our fitting approach for $t > 264$ d. In particular, no previous analysis has independently constrained the width at $t\lesssim 1000$ d, as the width is below the beam size of the interferometer at these times. However, it has been demonstrated in other analyses that modeling can constrain ring width well below the resolution of the interferometer, provided the diameter of the compact structure is sufficiently large \citep[see e.g.,][]{EHT1}. Indeed, we find that we can constrain the width of the SN 1993J ejecta so long as the ejecta are large enough that both the first null and second peak (first peak at $\rho=0$) are present in the visibility amplitudes, as shown in \autoref{fig:width_amp_demo}. This first occurs around $t\sim 300$ d. Though we do not fit the visibility amplitudes (instead favoring the log closure amplitudes for the same information), the visibility amplitudes help visualize and confirm that at earlier times ($t\lesssim 264$ d), the ejecta structure is effectively degenerate with a Gaussian or filled disk and the width is not constrainable without further assumptions. Thus, to maximize compatibility with our independent width modeling scheme, we do not fit the width and diameter independently earlier than $t=264$ d. However, we provide estimates of the ejecta size and brightness asymmetry at earlier epochs by fixing the prior of the width to $20\%$ of the diameter.

Outside of intentionally excluded days (see above), we also do not include results of fits performed on days 390, 451, 582, 873, or 1356. On these days, our modeling procedure produced clearly erroneous parameter estimates (e.g., infinitesimal widths) with complex marginal posterior distributions. We individually inspected these days and found they tended to correspond to poorer quality observations (e.g., worse coverage, unusually noisy data, etc.) than the successful epochs. Additional processing, akin to what was performed in e.g., \cite{Bartel2002}, \cite{Marcaide2009}, or \cite{Vidal2024}, may be required on these days, which will be explored in future work. However, in the interest of examining the results of our simple proof-of-concept procedure, we exclude these days that may require further data cleaning or processing steps.

\subsection{Fitting procedure and parameter estimation}
\label{sub:fitting_procedure_and_parameter_estimation}

For a given dataset and model prescription, we identify a ``best-fit'' model by maximizing a standard Gaussian likelihood function, which \cite{Chael2018} established as sufficient for closure phases and log closure amplitudes. The maximization and subsequent posterior exploration is performed via a Markov-Chain Monte Carlo (MCMC) procedure. We utilize the \texttt{python} package \texttt{emcee} in order to perform the MCMC fitting procedure \citep{emcee}. \texttt{emcee} is a Bayesian parameter estimator that implements the affine invariant MCMC ensemble sampler of \cite{MCMC}. We implemented \texttt{emcee} using uniform priors for all parameters. For parameters which are naturally bounded (e.g., $\arg \beta\in[0, 2\pi)$, $f\in[0, 1]$), the prior was bounded by the full range of the parameter. However, for parameters which are not naturally bounded, conservative (i.e., range-maximizing) parameter ranges were chosen. The diameter of the model $d$ and the width $\alpha$ are allowed a motivated lower bound of zero and an upper bound several times larger than the largest ejecta size reported in the literature. The fractional filling floor parameter $f$ was difficult to constrain due to degeneracies between $f$, $d$, and $\alpha$; however, we found $f\sim0.3$ was preferred on average throughout the observation. To simplify the posterior exploration, we fixed $f=0.3$ for all fits.

Beyond debiasing (see \autoref{sub:construction_of_closure_quantities}), no processing was done to the data prior to fitting. Additionally, a ``bootstrapping'' procedure was implemented prior to each science fit, wherein rapid fits are performed on simpler models to provide a meaningful starting point for the full complex model fit. For the \mring models, a fit to a simple symmetric ring model was performed first, and the best-fit parameters of this fit were used to initialize the full \mring fit. In testing, we found that the final fit solution was not sensitive to the starting point, but a meaningful initialization improved convergence times and provided more symmetric uncertainties. The bootstrapping fits were performed via a gradient-descent optimization routine. The final fits were performed via an MCMC procedure as described above, which used $10^3$ walkers and was allowed to run as long as required to achieve convergence. Convergence was assessed based on the improved Gelman-Rubin statistic \citep{ImprovedRhat} and a check that the chain length for each parameter was at least $50$x longer than the corresponding autocorrelation timescale \citep{emcee}. This strategy resulted in unimodal Gaussian posterior distributions for all parameters. Due to the simplicity of the posterior distributions, we computed confidence intervals around the maximum a posteriori parameter values using the highest posterior density interval \citep{HPDI}. 

\begin{figure*}
    \centering
    \includegraphics[scale=0.7]{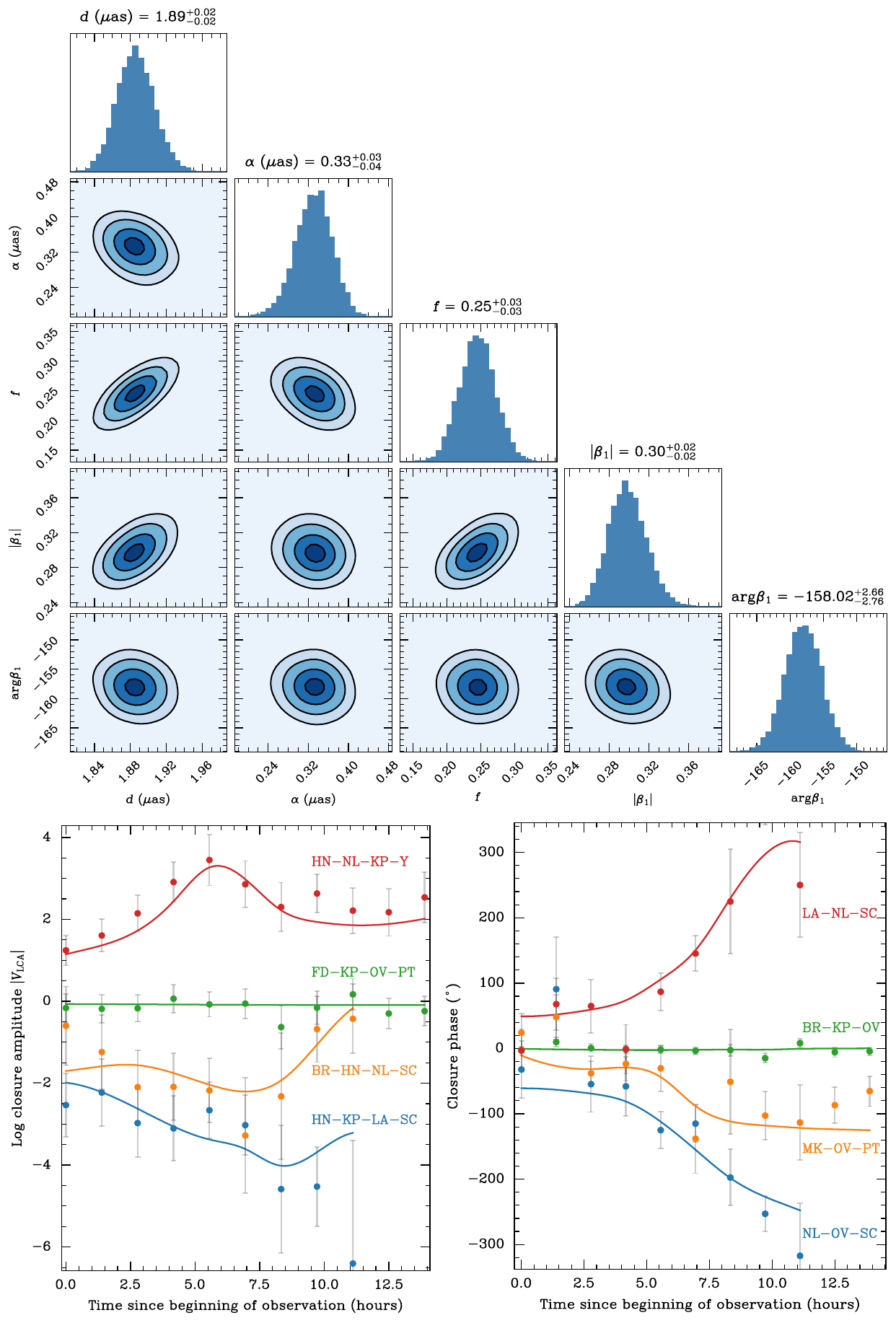}
    \caption{Here, we present a representative fit of an $m=1$ ring to the data from day 520. \textit{(top)} The corner plot resulting from the fit. All parameters have symmetric Gaussian marginal distributions, with mild correlations between the width, diameter, and fractional floor. \textit{(bottom left)} The maximum likelihood solution identified by the fit (solid lines) compared to the log closure amplitudes of the data (points) on four example quadrangles (HN-NL-KP-Y, FD-KP-OV-PT, BR-HN-NL-SC, HN-KP-LA-SC). The fit is a convincing representation of the data. The data have been averaged in time for clarity, but fits were performed on non-averaged data. \textit{(bottom right)} The maximum likelihood solution identified by the fit (solid lines) compared to the closure phases of the data (points) on four example triangles (LA-NL-SC, BR-KP-OV, MK-OV-PT, NL-OV-SC). The fit is a convincing representation of the data. The data have been averaged in time for clarity, but fits were performed on non-averaged data. }
    \label{fig:model_v_data}
\end{figure*}

An example fit (for day 520) is shown in \autoref{fig:model_v_data}, alongside model-data comparisons on sample closure triangles and quadrangles. The fit selected here is performed varying $f$, to demonstrate the preference for $f\sim0.3$. The fit demonstrates good convergence, with all parameters exhibiting symmetric Gaussian marginal posterior distributions. There are mild correlations between several parameters and the fractional floor. The modeled closure quantities show significant agreement with the observed closure quantities, regardless of choice of quadrangle or triangle. The model is also able to clearly reproduce trivial triangles and quadrangles, an important sanity check \citep{Chael2018}. We emphasize that this (the fit represented in \autoref{fig:model_v_data}) represents a meaningful width constraint $\approx250$ days earlier than any previous work; the earliest width constraints in our sample are $\gtrsim500$ days earlier than any previous analysis. Additionally, the angle of brightness asymmetry in the shell is constrained to better than $\lesssim40^\circ$, representing a novel statistically quantified constraint on the angular brightness distribution of the ejecta. We emphasize that interpretation of the recovered brightness asymmetry angle is model-dependent as the $\beta_1$ parameter only provides information on the location of the weighted mean of the azimuthal modulations, and not on the location of any particular hotspot or structure.

\section{Application to SN 1993J}
\label{sec:application_to_sn_1993j}

\begin{table*}[!ht]
\centering
\begin{tabular}{cccccccc}
\toprule
\textbf{Phase} (days) & $d$ (mas) & $\alpha$ (mas) & $\arg\beta_1$ ($^\circ$) & $|\beta_1|$ & $R_\mathrm{out}$ (mas) & $R_\mathrm{out}/R_\mathrm{in}$ & $\chi^2$ \\
\midrule
175  & $0.721 \pm 0.010$ & 0.1442${}^a$ & $109 \pm 5$   & $0.39 \pm 0.05$  & $0.460 \pm 0.082$ & $1.31 \pm 0.195$ & 1.310 \\
223  & $0.897 \pm 0.011$ & 0.1794${}^a$ & $126.1 \pm 0.5$ & $0.2720 \pm 0.0023$ & $0.561 \pm 0.091$ & $1.31 \pm 0.214$ & 1.820 \\
264  & $1.051 \pm 0.016$ & $0.18 \pm 0.05$ & $115 \pm 5$   & $0.283 \pm 0.022$ & $0.655 \pm 0.132$ & $1.287 \pm 0.129$ & 0.508 \\
306  & $1.192 \pm 0.013$ & $0.244 \pm 0.013$ & $131 \pm 3$   & $0.417 \pm 0.021$ & $0.763 \pm 0.107$ & $1.340 \pm 0.035$ & 1.309 \\
352  & $1.333 \pm 0.004$ & $0.229 \pm 0.024$ & $135.1 \pm 1.4$ & $0.242 \pm 0.008$ & $0.832 \pm 0.033$ & $1.304 \pm 0.067$ & 1.237 \\
520  & $1.888 \pm 0.009$ & $0.33 \pm 0.03$ & $158 \pm 3$   & $0.299 \pm 0.014$ & $1.178 \pm 0.074$ & $1.303 \pm 0.094$ & 1.276 \\
635  & $2.193 \pm 0.004$ & $0.366 \pm 0.021$ & $162.9 \pm 2.1$ & $0.142 \pm 0.006$ & $1.364 \pm 0.033$ & $1.265 \pm 0.054$ & 1.010 \\
686  & $2.361 \pm 0.025$ & $0.353 \pm 0.014$ & $192 \pm 14$  & $0.22 \pm 0.03$  & $1.435 \pm 0.206$ & $1.227 \pm 0.029$ & 1.402 \\
774  & $2.558 \pm 0.024$ & $0.59 \pm 0.05$ & $168 \pm 34$  & $0.337 \pm 0.022$ & $1.678 \pm 0.198$ & $1.418 \pm 0.172$ & 0.698 \\
997  & $3.135 \pm 0.005$ & $0.556 \pm 0.016$ & $230 \pm 3$   & $0.110 \pm 0.006$ & $1.958 \pm 0.041$ & $1.305 \pm 0.051$ & 1.121 \\
1107 & $3.407 \pm 0.009$ & $0.726 \pm 0.013$ & $275 \pm 5$   & $0.104 \pm 0.010$ & $2.229 \pm 0.074$ & $1.416 \pm 0.059$ & 1.541 \\
1253 & $3.828 \pm 0.013$ & $0.65 \pm 0.03$ & $254 \pm 8$   & $0.123 \pm 0.010$ & $2.384 \pm 0.107$ & $1.277 \pm 0.093$ & 1.357 \\
1532 & $4.398 \pm 0.020$ & $0.89 \pm 0.03$ & $514 \pm 3$   & $0.118 \pm 0.013$ & $2.812 \pm 0.165$ & $1.336 \pm 0.084$ & 1.468 \\
1693 & $4.675 \pm 0.007$ & $0.669 \pm 0.015$ & $281 \pm 4$   & $0.088 \pm 0.009$ & $2.783 \pm 0.058$ & $1.202 \pm 0.036$ & 1.334 \\

\bottomrule
\end{tabular}
\centering \\
\vspace{0.2cm}$^{a}$These days are fit using an $\alpha$ fixed to $0.2d$. 
\caption{Best-fit numerical values as a function of phase in days for radial expansion and asymmetry. The values represent the maximum likelihood solution obtained during the MCMC fitting performed as described in \autoref{sec:mring_modeling_and_fitting_framework}. The phase shown is relative to the explosion date of MJD 49074. The $R_{\textrm{out}}$ and $R_{\textrm{in}}$ correspond to the best-fit spherical shell model computed using the method described in \autoref{sec:width}. The $\chi^2$ values are normalized to the number of degrees of freedom present in the data, calculated using the uncertainty estimation method described in \autoref{sub:observations_and_data_reduction}. As discussed in \autoref{sub:observations_and_data_reduction}, our errors on the closure quantities are likely significantly underestimated due to the low $S/N$ of the data, and this is reflected in the unusually low parameter uncertainty estimates reported in this table. Therefore, for the remaining analysis, we scale the uncertainties such that the reduced $\chi^2$ of a reasonable model (e.g., power-law for the diameter) is $\approx1$. These scalings are specific to each parameter and are discussed in the relevant sections.} 
\label{tab:results}
\end{table*}

\subsection{Radial evolution of the ejecta}
\label{sub:radial_evolution_of_the_ejecta}

\subsubsection{Evolution of the ejecta size}
\label{sub:evolution_of_the_ejecta_size}

\begin{figure}
    \centering
    \hspace*{-1cm}\includegraphics[scale=0.55]{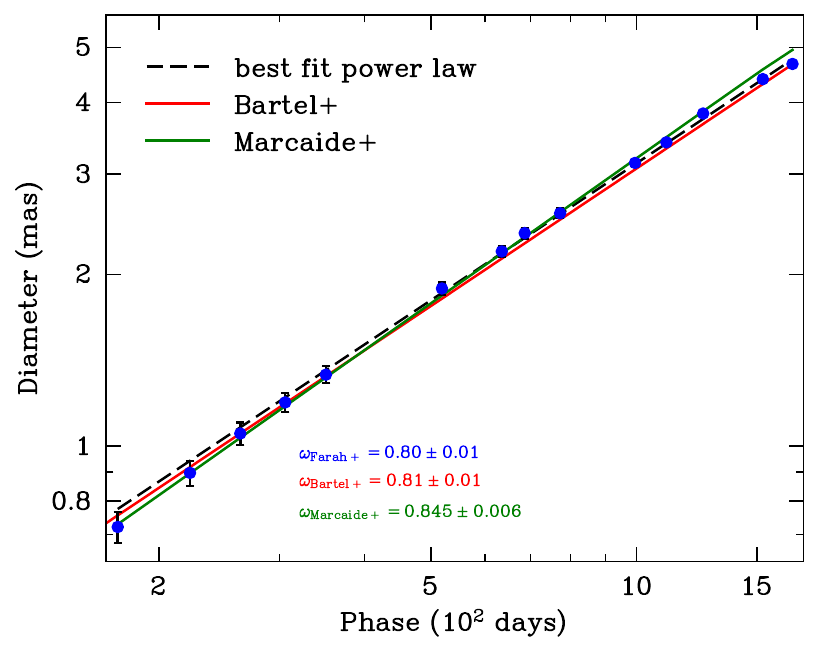}
    \hspace*{-1.15cm}\includegraphics[scale=0.55]{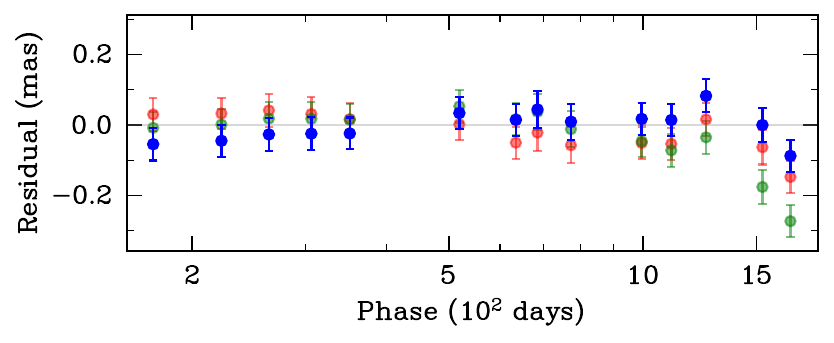}
    \caption{\textit{(top)} Radial expansion recovered via our modeling approach. The expansion is well-approximated by the power-law with $\omega\approx0.80\pm0.01$ shown in \autoref{fitting:eq:power_law} (black line). 
    We include a comparison with \cite{Marcaide2009} (green line) and \cite{Bartel2002} (red line, re-fit using one-component power law) to demonstrate that our radial expansion is in agreement with the recovered value in the literature. We report comparable agreement with both \cite{Marcaide2009} and \cite{Bartel2002} at all epochs.  \textit{(bottom)} Residual comparison of deceleration reported by this work, \cite{Bartel2002}, and \cite{Marcaide2009}. The magnitude of the deviation from both the power-law fit and the literature values is $\lesssim0.2$ mas at all epochs. Uncertainties displayed and fit are scaled so that the reduced $\chi^2$ of the power-law fit is $\approx 1$.}
    \label{fig:diam_powerlaws}
\end{figure}

We report monotonic radial expansion in the ejecta. The modeled ejecta diameter for each epoch, along with the associated uncertainties, are shown in \autoref{fig:diam_powerlaws}. The best-fit values for each epoch and the resulting $\chi^2$ are shown in \autoref{tab:results}. The values in this table represent the maximum likelihood solution obtained during the MCMC fitting performed as described in \autoref{sec:mring_modeling_and_fitting_framework}. 
The supernova phase reported in the table is computed relative to the explosion date of March 28, 1993 (MJD 49074). The normalized $\chi^2$ values are derived from the likelihood of the maximum likelihood solution and are normalized to the number of degrees of freedom present in the data.

The size measurement was largely constrained by data on baseline lengths comparable to the location of the first null in the amplitudes, which is the strongest proxy available for the overall size of the compact emitting region. The location of this null $\rho_{\textrm{null}}$ increasingly tends towards shorter baselines as the angular size of the ejecta $d$ increases, following $\rho_{\textrm{null}}\propto 1/d$.  The magnitude of the uncertainty in the measurement in a noisy dataset is therefore affected by the presence of numerous and high-$S/N$ baselines with lengths at $\approx \rho_{\textrm{null}}$. Since the $S/N$ ratio does not vary monotonically with baseline length, the uncertainties on the diameter were similarly inconsistent. 
The uncertainties on the diameter are significantly underestimated, as a simple power-law fit produces $\chi^2 \gg 1$. The underestimation is likely due to the reasons discussed in \autoref{sec:data}. To mitigate this, we scale the uncertainties so that the $\chi^2$ of a power-law fit is $\approx 1$. These scaled uncertainties are shown in \autoref{fig:diam_powerlaws}, and the power-law fit is discussed below.

The expansion as a function of supernova phase is shown in \autoref{fig:diam_powerlaws}. The expansion is consistent with a power-law of the form
\begin{equation}
    d(t) = R_0\left(\frac{t}{\tau_0}\right)^{\omega}.
\label{fitting:eq:power_law}
\end{equation}
with $R_0 = 2.48 \pm 0.18$ mas, $\tau_0 = 750 \pm 39$ days, and $\omega = 0.80 \pm 0.01$. This power-law is compared to our modeled radial expansion as well as the results at 8 GHz from \cite{Marcaide2009} and \cite{Bartel2002} in \autoref{fig:diam_powerlaws}. The literature results are scaled to account for a difference between the definition of ``size''--our model reports a size corresponding to a midpoint with a flux falloff on either side, whereas previous literature defined the size using the edge of the compact emitting region (see further discussion in \autoref{sec:width}). 

The expansion we obtain is grossly consistent with that reported in the literature. Our deceleration parameter $\omega = 0.80 \pm 0.01$ is consistent to $\sim1\sigma$ with the value of $\omega(t<\textrm{1500 days}) = 0.845 \pm 0.005$ reported in \cite{Marcaide2009} for the same epochs, as well as the value of $\omega\approx0.80$ reported in the more recent \cite{Vidal2024}. For the results presented in \cite{Bartel2002}, we re-fit the shell sizes for the range of our analyzed epochs using our power law model presented in \autoref{fitting:eq:power_law} and find $\omega=0.81\pm0.01$, consistent with our value up to $1\sigma$. 


\subsubsection{Evolution of the ejecta width}
\label{sec:width}

\begin{figure}
    \centering
    \includegraphics[scale=0.55]{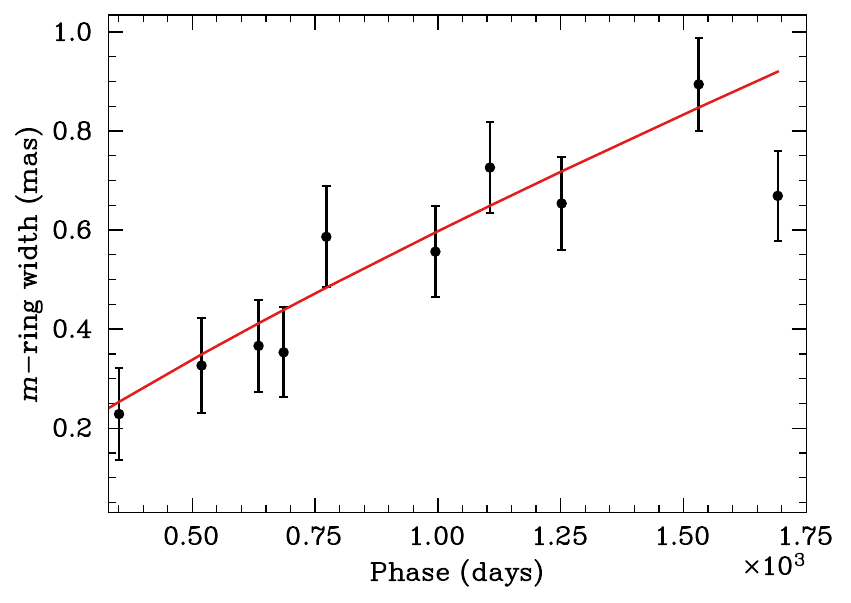}
    \includegraphics[scale=0.55]{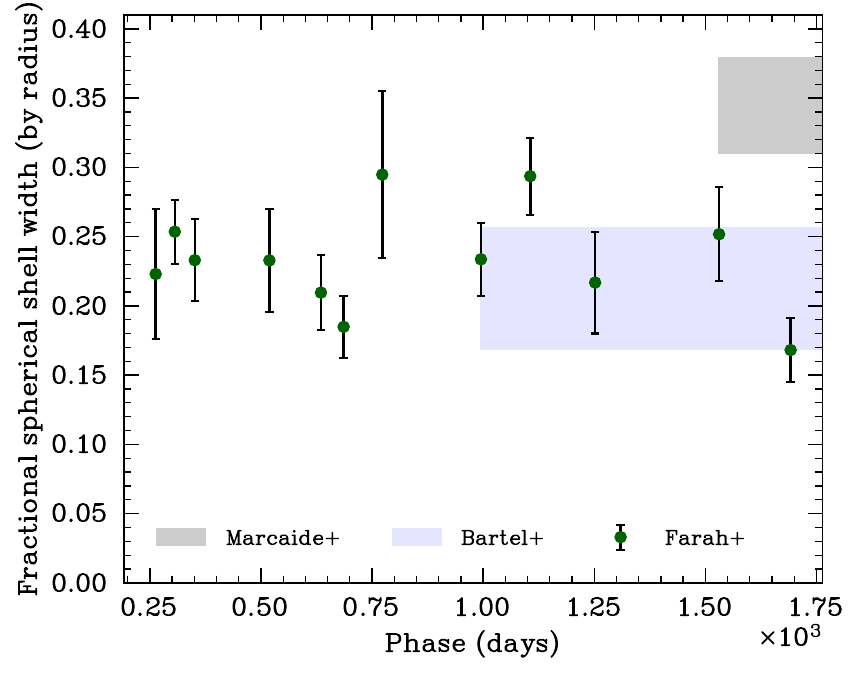}
    \caption{(\textit{top}) Evolution of the ejecta width ($\alpha$ parameter in the \mring) as recovered by our model. Displayed are the raw width values recovered by the model. The width evolution is well-approximated by 20\% of the radial expansion (shown by the solid red line). The modeled width evolution tracks the consensus diameter evolution closely. (\textit{bottom}) Comparison of transformed (see \autoref{sec:width}) widths between literature model fits (gray and blue shaded regions) and our analysis (green points). For the comparison to \cite{Bartel2002}, we computed a mean and standard deviation of the widths reported in those analyses on epochs overlapping with our analysis (namely, $t \gtrsim900$ d), visualized as the blue band. For the comparison to \cite{Marcaide2009}, we quote the spread of values they report, beginning around $t\sim1500$ d (gray band). We globally recover widths consistent $\lesssim1\sigma$ with the analysis of \cite{Bartel2002}. By contrast, our widths are almost universally inconsistent with the results of \cite{Marcaide2009}, being $\approx30\%$ smaller on average for all epochs. More recently, \cite{Vidal2024} found a fractional width of $\approx0.31\pm0.08$, which is more consistent with our analysis. Uncertainties displayed are scaled so that the reduced $\chi^2$ of the power-law fit is $\approx 1$.}
    \label{fig:width_evolution}
\end{figure}

Our modeling approach recovers a largely monotonically increasing width, ranging between an angular size of $\approx0.2$ mas at the earliest epochs to $\approx0.8$ mas at the latest, corresponding to a ${\approx}4$x increase over the first $\approx1.5\times10^3$ days of the supernova ejecta expansion. The evolution of the width is shown in \autoref{fig:width_evolution}. The width evolves in a manner that can be approximated by 20\% of the power-law radial expansion solution reported in \autoref{sub:radial_evolution_of_the_ejecta}. 

Similarly to the diameter, the parameter uncertainty appear to inaccurately represent the variability in the model fit estimates, which represent unphysical evolution if taken at face value. We rescale the uncertainties on $\alpha$ and derived quantities to achieve a reduced $\chi^2 \approx 1$ on a power-law model fit. The $\alpha$ evolution and rescaled uncertainties are shown in \autoref{fig:width_evolution}.

Due to differences in model construction, it is generally not possible to compare model parameter outputs to other models or images directly. For example, a spherical shell model was previously used in the literature to estimate the width of the SN 1993J ejecta \citep{Bartel2002,Marcaide2009}. The width in this model corresponds to the true width of the emission region, assuming the ejecta is indeed a spherical shell of uniform emissivity. However, this width cannot be naively compared to our model, which measures the projected width of the spherical shell in the two-dimensional observer plane. 

\begin{figure*}
    \centering
    \includegraphics[scale=0.64]{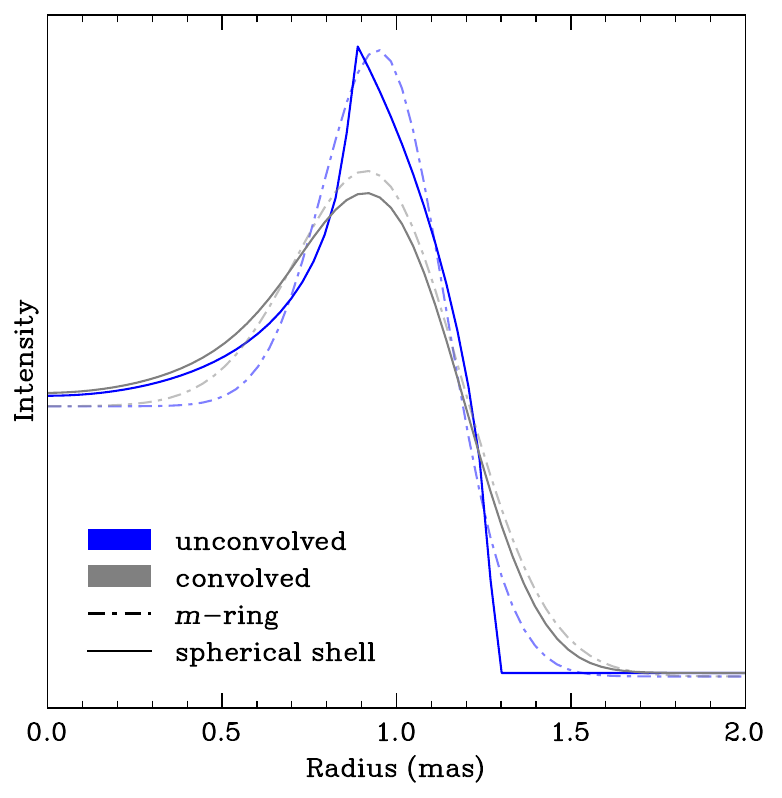}
    \includegraphics[scale=0.63]{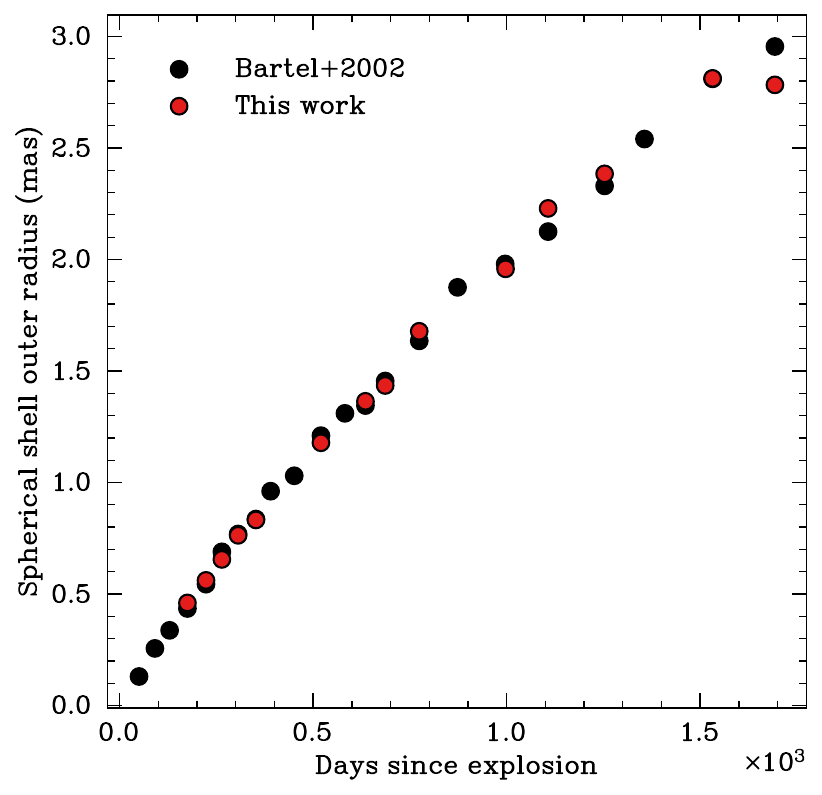}
    \caption{\textit{(left)} Comparison of the radial profile of the $m$-ring model (dot-dashed line, $f=0.4, d=2$ mas, $\alpha=0.4$ mas) to the best-fitting spherical shell model (solid line) used in e.g., \cite{Bartel2002,Marcaide2009}. The models feature different parameterizations but largely describe similar profiles. Even without convolution (blue), the $m$-ring model is an acceptable approximation to the spherical shell. Under convolution with an $\approx1$ mas beam, the models become nearly indistinguishable. Note that the models favored a fractional floor parameter ($f\approx 0.3$) lower than displayed here.\textit{(right)} Comparison of the \cite{Bartel2002} best-fit spherical shell outer radius to the $m$-ring size parameter transformed according to the procedure in \autoref{sec:width}. The consistency with the results from \cite{Bartel2002} (which fit a spherical shell model directly) is an important sanity check for our procedure. Errors for the \cite{Bartel2002} diameter are $\approx1/4$ the radius of the circles at early times, growing to almost a full radius at late times. By contrast, the errors on $R_{\textrm{out}}$ from our analysis tended to be large ($\sim10-20\%$ of the radius) at early times and smaller ($\sim5-10\%$) of the radius at late times.}
    \label{fig:sshell_vs_mring}
\end{figure*}

In order to more meaningfully compare our width results with those in the literature, we transform our \mring width $\alpha$ in terms of the spherical shell width $\theta_{w}$ by finding a best-fitting spherical shell radial profile for any given $m$-ring. We demonstrate this procedure in \autoref{fig:sshell_vs_mring}. The \mring and spherical shell models have similar radial profiles, despite the \mring not being physically motivated. Under convolution with the nominal beam (particularly at early epochs when the ejecta are small), this similarity makes them almost indistinguishable. Given an \mring profile with parameters $d$, $\alpha$, and $f$ identified by our modeling scheme, we can identify the equivalent spherical shell parameters $R_{\textrm{out}}$ and $R_{\textrm{in}}$ by fitting the spherical shell radial profile to the \mring radial profile. This transformation yields values for $R_{\textrm{out}}$ that are similar to best-fit values from the literature (see \autoref{fig:sshell_vs_mring}), validating the transformation procedure. The radius and width of the best-fitting spherical shell are used to construct a transformed fractional width, which we compare to literature values in \autoref{fig:width_evolution}. 

For the comparison with \cite{Bartel2002} and \cite{Marcaide2009}, we determined the mean and standard deviation of the widths given in those studies for epochs overlapping with ours (specifically, $t \gtrsim 900$ d for \citealt{Bartel2002} and $t \gtrsim 1500$ d for \citealt{Marcaide2009}). The results of our comparison are visualized in \autoref{fig:width_evolution}. For the later epochs, \cite{Marcaide2009} reported a spherical shell model width ranging from $31\pm2$\% to $37.8\pm1.3$\%.
This is in substantial disagreement with both the early-time results reported by our analysis and \cite{Bartel2002}. By contrast, we find agreement up to $1\sigma$ with the results of \cite{Bartel2002}for most days. The fractional shell width eventually increased after $\sim6$ years \citep{Bietenholz2010}, which is not probed by our analysis.  


\subsection{Evolution of the brightness asymmetry}
\label{sub:evolution_of_the_brightness_asymmetry}

We present a unique measurement of the asymmetry of the SN 1993J ejecta---to wit, the first measurements with statistically meaningful uncertainties. To assess the asymmetry of the ejecta, we fit the \mring model with $m=1$ modes to all epochs in our sample. We report an evolving asymmetry in the azimuthal structure of the ejecta with a southern bias. 

\begin{figure*}
    \centering
    \includegraphics[scale=0.3]{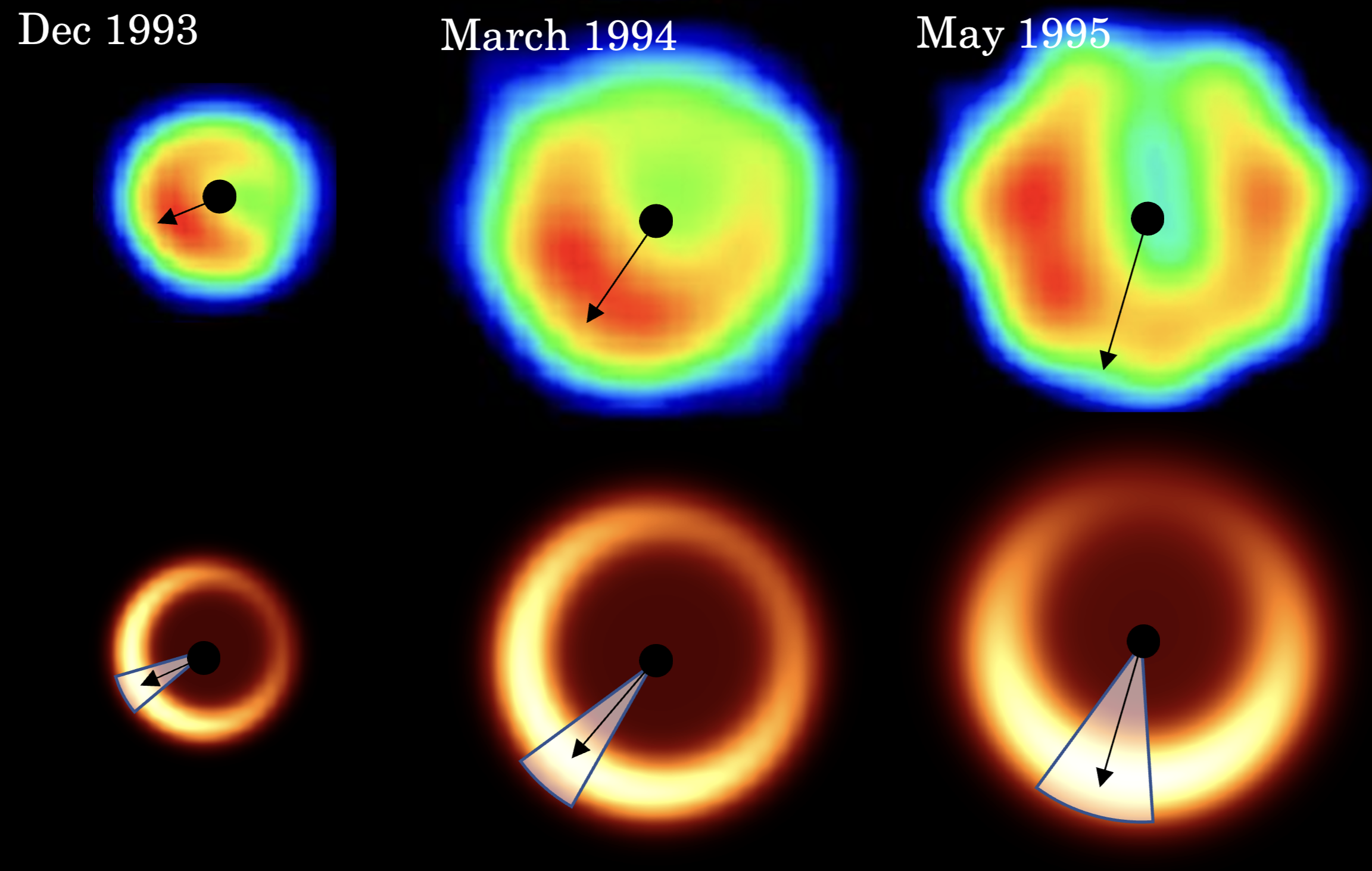}
    \hspace*{-0.8cm}\includegraphics[scale=0.72]{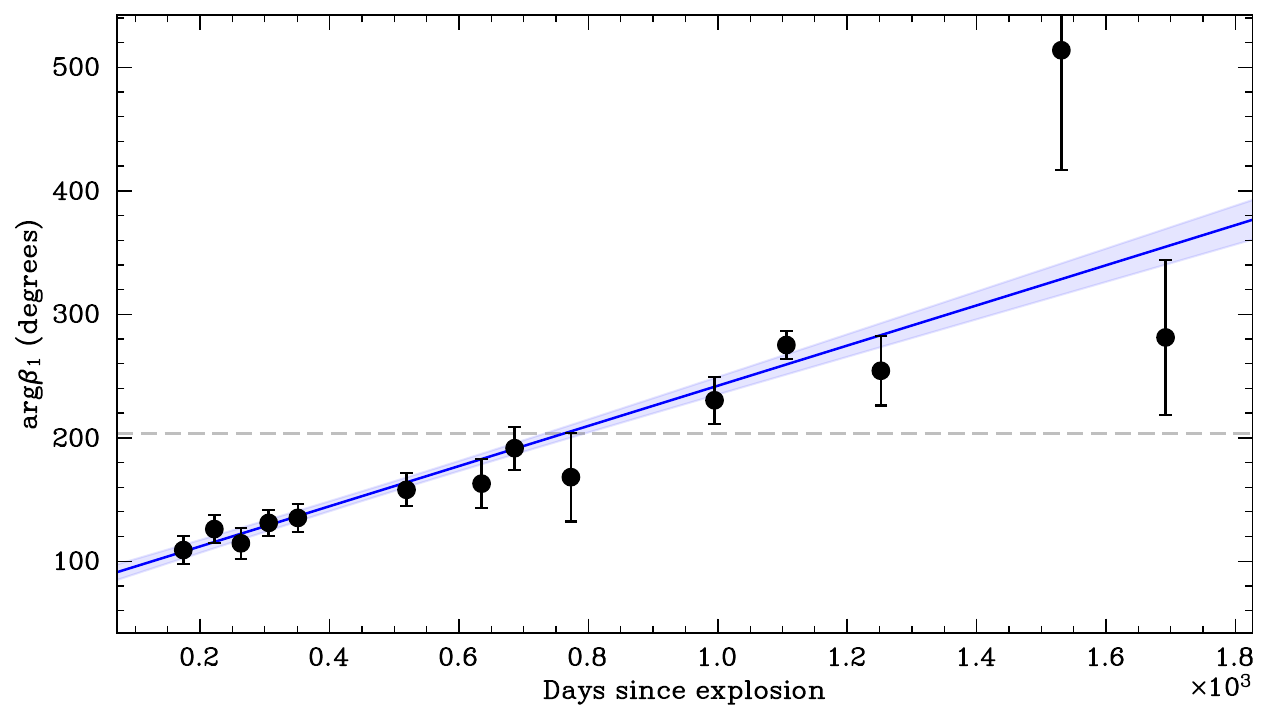}
    \caption{\textit{(top)} Visual comparison of the best-fit \mring models (afmhot colormap) to images from \cite{Bietenholz2003} (jet colormap). Black arrows indicate the approximate mean flux asymmetry angle ($\arg\beta_1$ in the \mring) in each image. The partially transparent fan shows a $\approx30^\circ$ uncertainty in the angle recovered by the \mring. For simpler azimuthal brightness profiles, this angle corresponds to the peak intensity angle, which is well-modeled by the \mring. However, as the ejecta become more complex, the \mring $\arg\beta_1$ parameter can no longer be interpreted as the location of a particular hotspot. \textit{(bottom)} Evolution of the circular mean asymmetry as a function of time (black points). We find that on average the azimuthal asymmetry drifts approximately $0.16^\circ \pm 0.02^\circ$ per day counterclockwise (i.e., from East to West, blue dashed line), providing evidence of a drift in the ejecta brightness distribution. The uncertainties are scaled following the procedure described in \autoref{sub:evolution_of_the_brightness_asymmetry} to achieve $\chi^2\approx 1$.}
    \label{fig:asymmetry}
\end{figure*}

The asymmetry in the \mring model is generated by the $\beta$ parameters. For each $\beta$ contribution, the modulus $|\beta|$ sets the magnitude of the deviation from uniform (i.e., higher $|\beta|$ corresponds to a stronger cosine mode), while the argument $\arg\beta$ sets the direction. Under the construction of our model, $\arg\beta = 0^\circ$ corresponds to the north (up) of the map, and $\arg\beta = 90^\circ = \pi/2$ corresponds to the east (left). For simple ejecta structure with one minimum and one maximum and a largely symmetric azimuthal brightness profile, the $\arg\beta_1$ parameter approximates the location of the maximum. However, for more complex azimuthal brightness profiles with multiple maxima (e.g., for $t > 996$ d), the $m=1$ proxies the location of the circular mean asymmetry and not necessarily the location of any particular maximum.

To assess the asymmetry, we draw samples of the complex $\beta$ parameters from the posterior explored during the fit and measure an average asymmetry position from each using the method of circular mean \citep{Buss2001}. The circular mean provides an estimate of asymmetry that is agnostic to choice of zero angle direction in the polar coordinate system. This distribution of asymmetry positions contributes to the mean (and standard deviation) which becomes our measurement (and model uncertainty) for each epoch. In general, we found that our uncertainties were significantly underestimated, as anticipated (see \autoref{sec:data}). Therefore, we scale the uncertainties such that the reduced $\chi^2$ of a linear fit is $\approx 1$. The deviation from linear tended to increase as the ejecta structure became more complex and symmetric (on average), which we measured using the $|\beta_1|$ parameter. At early times, when the $|\beta_1|$ was large (indicating the domination of a single cosine mode, confirmed by the images), we do not find a significant deviation from the linear drift behavior. However, in the last few epochs, as the $|\beta_1|$ becomes consistent with zero (i.e., no brightness asymmetry on average, and therefore an asymmetry ``direction'' is less meaningful) the deviation is significant. To mitigate this, we allowed a more aggressive scaling of the uncertainties of $\arg\beta_1$ measurements on days with very low $|\beta_1|$. The evolution of the $\arg\beta_1$ parameter, processed using the above procedure, is shown in \autoref{fig:asymmetry}. We find a bulk asymmetry toward the south centered around $\approx 203^\circ$, measured counterclockwise from north, averaged across all epochs.  The data are relatively well-described by a linear drift model with $d\arg\beta_1/dt = 0.16^\circ\pm0.02^\circ$ per day counterclockwise (i.e., from East to West). The estimated linear drift is inconsistent with no drift to $\gg 5\sigma$, providing evidence of a drift in the ejecta brightness distribution.

We compare and contrast the imaging and modeling approaches to recovery of the azimuthal brightness modulations in the top panel of \autoref{fig:asymmetry}. The images from \cite{Bietenholz2003} show a clearly evolving shell structure, resembling a horseshoe pattern rotating counter clockwise during the first several hundred days. At early times, the ejecta structure is fairly simple, with one peak brightness location approximately $180^\circ$ opposite from the least bright point along the ring. This simple modulation profile is well-approximated by the $m=1$ \mring, which produces a similar structure via the cosine modulation. Under these conditions, the $\arg\beta_1$ parameter of the \mring model aligns with the peak brightness location in the images, which for simple structure is almost identical to the circular mean brightness asymmetry angle (the quantity that $\arg\beta_1$ actually probes). However, at later times, the azimuthal behavior of the ejecta becomes more complex, and as a result the \mring is a poorer approximation. In the May 1995 example shown in \autoref{fig:asymmetry}, an additional hotspot has formed opposite the peak brightness location, moving the mean asymmetry angle partway between the two hotspots. The \mring captures this movement of the mean asymmetry angle accurately but does not provide full information about the ejecta structure. The primary benefit of the \mring approach, then, comes from the information afforded about the uncertainty in the best-fit mean asymmetry angle; i.e., the range of mean asymmetry angles plausibly supported by the data. While the imaging approaches can reproduce arbitrarily complex brightness profiles, there is no information available about the space of image structures that fit the data with similar success.

Previously, \cite{Heywood2009} proposed that this rotation could be an artifact caused by the $(u, v)$-coverage evolving with time. Modeling approaches are insensitive to coverage artifacts (the missing information due to the coverage is incorporated into the uncertainties), and the brightness asymmetry uncertainties produced by our analysis allow us to investigate the robustness of a drift or rotation. The previous linear fit provides evidence of a drift throughout the observation as a whole, supporting the finding of previous imaging-based analyses.

\subsection{Uncertainties and systematic error}
\label{sec:sys_error}

The reported uncertainties associated with quantities measured via our \mring model fitting approach are typically inconsistent with those reported for similar quantities in the literature, likely due at least in part to procedural factors discussed in \autoref{sec:data}.
By contrast, the uncertainties reported by our analysis on the azimuthal brightness variation are novel and have no established result available for comparison. Here, we examine possible sources of systematic error and evaluate the impact on our results.




\subsubsection{Absorption in the center of the radio shell}

\cite{Bartel2002} investigated the impact of the opacity of the radio shell (particularly at early times) on estimates of the outer radius. The spherical shell model used to perform measurements did not have an absorption component, leading to a bias of up to $\approx10$\% on the size at the earliest days. However, this bias was not considered an issue for several reasons. First, the calculation was performed assuming the worst-case scenario of complete absorption, which \cite{Bietenholz2003} found was not the case. Second, the bias decreased with resolution, so the impact was primarily on the earliest days. Third and finally, the bias was in the opposite direction of the bias created by the lack of angular variability in the chosen model, so the two systematic biases were presumed to cancel. We cannot rely on these motivations---our analysis focuses on early data (when the opacity effects may be significant) and as discussed in \autoref{sub:complexity_of_the_mring}, the \mring model is significantly less affected by systematic bias to azimuthal complexity. However, our analysis is indeed largely unaffected by this source of systematic error, as the fractional floor parameter $f$ of the \mring acts identically to an opacity parameter, and therefore we can account for varying absorption at both early and late epochs.

\subsubsection{Size-width degeneracy}

There exists a minor degeneracy between angular size and width, discussed by both \cite{Bartel2002} and \cite{Marcaide2009}. \cite{Bartel2002} noted that at an assumed unresolved shell width of $\approx25$\% (corresponding to an \mring $\alpha\approx34$\%), the resulting bias in shell size at early epochs was $\approx2$\%, and fixed the shell width at the earliest epochs ($t<$ day 900).  \cite{Marcaide2009}  reported that, due to this degeneracy, simultaneous fits of angular size and width were difficult to obtain without strong correlations and incorrect parameter recovery.
To counter this, \cite{Marcaide2009} fixed the model width, fit the angular size, then fixed the angular size and re-fit the width. This approach does not constitute an independent measure of size or width, especially given the bias in width of up to 25\% reported by \cite{Marcaide2009} when the diameter is incorrectly estimated by just 5\%. 

Our analysis is largely unaffected by this source of systematic error, as we fit the size and width simultaneously for epochs $t \geq 264 $ days. As a result, we do not need to fix the width for these epochs as in \cite{Bartel2002} (which fixed the width between $t=389$ and $t=996$ days, but fit them simultaneously after) or fit width and size separately as in \cite{Marcaide2009} (which fixed the width and fit the diameter, then fixed the diameter to the best-fit diameter and re-fit the width). Further, we do not reproduce the difficulties reported by \cite{Marcaide2009}, and recover clean, uncorrelated posteriors for simultaneous fits of width and size, as shown in \autoref{fig:model_v_data}. Since we are fitting size and width simultaneously, any bias resulting from the size-width degeneracy will be incorporated into the uncertainty reported by the MCMC fitting algorithm. Our fits to days $t < 264$ d (which fixed the width to 20\% of the diameter) may be affected by this 2\% bias.


\subsubsection{Complexity of the \mring}
\label{sub:complexity_of_the_mring}

\cite{Bartel2002} investigated the impact of azimuthal modulation along the model ridge as a potential source of systematic error in their size and width measurements. Though their images reported in \cite{Bietenholz2001,Bietenholz2003,Bietenholz2005} showed angular variation in the brightness profile of the compact object, the spherical shell model used to conduct measurements was circularly symmetric in projection and was unable to reproduce these features. They calculated a systematic error corresponding to a $\approx5$\% reduction in the size of the outer radius, assuming a fixed width, at the earliest epochs, where the effect was most pronounced. The effect diminished as the modulation increased in complexity with time. 

A novel aspect of our analysis is the incorporation of azimuthal complexity via the introduction of \mring modes with $m>0$. As reported in \cite{Bartel2002}, the properties of an asymmetric source can be misrepresented if fit to a model without asymmetry. We investigated the impact of our \mring implementation on our size estimates by re-fitting the observations using a symmetric \mring model ($m_{1, 2, \ldots} = 0$) as well as fitting a synthetic observation of an $m$-ring with $m=3$ complexity. Across all epochs, we recover angular sizes for the $m=0$-only fits consistent with the $m=1$ fits to $\lesssim1\sigma$, with no evidence for a systematic offset or bias. However, we found that reducing the number of parameters improved the uncertainties. For example, on day 520, our full \mring analysis reports a size uncertainty of $\approx \pm1$\%. However, fitting with only $I_0$, $d$, $\alpha$, and $f$ results in an uncertainty $\sim2$x smaller, of about $0.5\%$. Increasing the number of $m$-ring modes from $m=0$ to $m=1$ results in an increased size uncertainty of $\approx1\%$. Finally, we found that fitting a model with $m=3$ complexity with the $m=1$ model results in an approximately $2\%$ reduction in the measured diameter (compared to the true diameter), which is consistent within the associated uncertainties. By contrast, there was $\lesssim1\%$ bias in the width measurement, which in the \mring model is not affected by the presence of asymmetry. We mitigate the azimuthal complexity bias by only fitting to days which have a significant $m=1$ mode contribution (even if those with significant $m=1$ mode contribution have more complex substructure) based on imaging \citep{Bietenholz2003,Marcaide2009}, as discussed in \autoref{sub:epoch_selection_criteria}.

\section{Discussion}
\label{sec:discussion}
The results obtained using our modeling approach yields a radial expansion that is largely consistent with the results produced by previous imaging analyses. As shown in \autoref{fig:diam_powerlaws}, the modeled radial expansion is consistent with results in the literature at all epochs. We find statistically greater consistency with the width evolution modeled in \cite{Bartel2002} and \cite{Vidal2024} versus \cite{Marcaide2009}. 

Our analysis fits size and width simultaneously \citep[in contrast to some previous analyses, e.g.][]{Marcaide2009} and fits width at all epochs \citep[in contrast to some previous analyses, e.g.][]{Bartel2002,Vidal2024}, which provides greater confidence in the size and width measurements.  Given that the size of the ejecta is the easiest quantity to recover in this type of analysis, our determination of a size and evolution consistent with previous analyses and simulation is a strong verification that our results are a proper description of the dataset.

As reported in \autoref{sub:evolution_of_the_brightness_asymmetry}, the asymmetric maps obtained by our modeling framework are somewhat consistent with the image reconstructions attempted on this dataset previously. We find a bulk asymmetry to the south of the image across all epochs, which decreases in magnitude in the later epochs as the ejecta expands. Our approach, however, offers a unique estimate of the uncertainties on the bulk asymmetry. The sparsity problem of image reconstruction makes it challenging to assess which reconstructed features are believable, and to what degree. Forward-modeling-based imaging approaches (e.g., regularized maximum likelihood) can approximate statistical robustness by performing feature-extraction on a ``top-set'' of best-fitting images, somewhat akin to a posterior exploration. The uncertainties associated with our best-fit angular profiles properly characterize the space of possible azimuthal characteristics the data represent, to the extent our model is able to describe the data. As a result, while reconstructions performed in the literature seem to indicate an asymmetry in the ejecta towards the south, we can uniquely characterize the amount, and assess its statistical significance. 

The ability to probe such azimuthal brightness variations directly from the data, and with well-characterized uncertainties, is critical for distinguishing real physical evolution from artifacts. \cite{Heywood2009} argued that the apparent rotation in imaging studies \citep[e.g.,][]{Bietenholz2003,Marti-Vidal2011} could arise from changing $(u,v)$-coverage; in contrast, our modeling incorporates coverage incompleteness into the posterior uncertainties, rendering the measurement insensitive to such effects. The persistence of the rotation signal in our fits, and its statistically significant two-phase behavior, argues against a coverage artifact and in favor of intrinsic evolution in the ejecta.

Our analysis confirms that the ejecta are persistently asymmetric toward the south, with a mean asymmetry angle of $\sim203^\circ$ (measured counterclockwise from north). A global linear fit yields a counterclockwise shift of $0.16^\circ \pm 0.02^\circ$ per day, inconsistent with zero at $\gg 5\sigma$. From a theoretical perspective, these measurements open a new window into the physics of young supernova remnants. Detecting and quantifying azimuthal brightness variations this early after explosion provides a direct test of models in which asymmetric density or wind profiles \citep[e.g.,][]{Toledo2014}, interaction with anisotropic circumstellar material \citep[e.g.,][]{Jun1996b}, or clumpy magnetic field amplification govern the morphology of the expanding shell \citep[e.g.,][]{Jun1996}. The evidence for a rapid early drift that slows at later times suggests a dynamical transition in the ejecta–environment coupling, for which models of shell interaction and brightening must account.

As discussed in \autoref{sec:application_to_sn_1993j}, we report an expanding projected ejecta width equal to approximately $0.22\pm0.01$ (weighted mean and uncertainty of all fit epochs) of the projected shell radius. In contrast to previous analyses, we can robustly fit the width of the radio feature to as early as $t\approx264$ days, $\sim700$ days earlier than estimates from the literature (e.g., \citealt{Bietenholz2003}, which fit the width down to 996 days). Our fractional width is statistically significantly lower than later results reported in \cite{Marcaide2009} (by a factor of $\approx1\textrm{--}2$). Further, our widths are statistically consistent to $\lesssim1\sigma$ with the results of \cite{Bartel2002} and \cite{Vidal2024}. There are several nuances about this comparison that should be emphasized.

First, the definition of ``width'' across model classes is not consistent, and, as a result, one model's ``width'' cannot be blindly compared to another's. The construction of width in the \mring model (convolution of an infinitesimally thin ring with a Gaussian kernel) is similar to how width is measured in image-domain feature extraction methods, but distinct to how the width is constructed in the spherical shell model of e.g., \cite{Bartel2002} and \cite{Marcaide2009}. We address this by identifying our \mring width with the spherical shell width which best reproduces the best-fit \mring radial profile.

Second, the \mring model is phenomenological and not as physically-motivated as models used in other analyses (e.g., the spherical shell of uniform emissivity). As a result, there is a valid concern that, if the true underlying shape of the SN 1993J ejecta is a non-\mring morphology, our result could be biased. This concern particularly applies to the width, which is sensitive to both data noise and model construction. We examined this concern in the case of the spherical shell of uniform emissivity, and found that: (a) for any spherical shell of uniform emissivity, there is at least one \mring configuration that can reproduce the visibility function up to the resolution of the interferometer, and (b) the ability of the \mring to produce Gaussians, disks, rings, and crescents (see \autoref{sec:mring_modeling_and_fitting_framework} for a demonstration) with a single component provides it more flexibility than almost any other model type. Finally, comparing the ability of the \mring model and a spherical shell of uniform emissivity to describe the data using a statistical $f$-test indicates that a spherical shell of uniform emissivity does not describe the data better on any epoch than the best-fitting \mring. 
Therefore, while caution needs to be taken in direct model-to-model comparisons, we are confident that the \mring model is sufficient to describe the data. 

\section{Conclusions}
\label{sec:conclusions}
The \mring\ model provides a streamlined yet flexible framework for describing the morphology of SN 1993J. Its analytic visibility function makes it computationally efficient for rigorous inference methods, while its parameterization can reproduce Gaussians, disks, rings, and crescents, spanning the circularly-symmetric morphologies expected in the data. This generality allows us to capture the projected spherical-shell geometry used in earlier analyses, while also enabling explicit characterization of azimuthal brightness variations. The construction is, however, phenomenological rather than physically motivated, and at later epochs—when the ejecta develop multiple hotspots and more complex substructure—the \mring\ provides only an approximation to the mean asymmetry rather than a complete description.

We mitigate these limitations by directly comparing the radial profile of the \mring\ to spherical-shell models and demonstrating that the \mring can reproduce the most important features of the spherical shell. We exploit this similarity to map \mring quantities to the equivalent (i.e., best-fitting) spherical shell, mitigating the unphysical construction of the \mring by translating to physically motivated quantities. Additionally, by fitting closure quantities rather than visibilities, our analysis is rather independent of previous approaches, despite relying on the same dataset. Closure phases and log closure amplitudes make our analysis insensitive to calibration systematics that have complicated prior analyses of SN 1993J. However, as a drawback, it is not possible with this method to calibrate the data in Jansky and obtain the total flux density. A key remaining limitation of our procedure is the apparent underestimation of parameter uncertainties. We investigate possible sources for the underestimation but rely on adjustment via the reduced $\chi^2$ metric to correct the underestimation.

We summarize our main results:
\begin{enumerate}
    \item \textbf{Size characterization of the expanding ejecta}. We measure the diameter from 175 d to 1693 d and find an average power-law index of $\omega=0.80\pm0.01$, consistent with previous analyses.
    \item \textbf{Early-time measurement of the width}. For the first time, we have presented geometric measurements of the ejecta width earlier than $\sim1000$ d. We independently fit the ejecta diameter and width down to $264$ d, $\sim700$ days than previous analyses. We find an average ejecta width of $0.22\pm0.01$ between $t=264$ d and $t=1693$ d. 
    \item \textbf{Azimuthal brightness modulation fit}. We identify a brightness pattern which rotates from east to south with a rate of $0.16^\circ \pm 0.02^\circ$ per day during first few hundred days, after which the azimuthal structure becomes more complex with increasing angular resolution.
\end{enumerate}
This analysis provides a proof-of-concept for interferometric closure quantity modeling of young radio supernovae and will be useful for future VLBI observations of explosive transients.

\acknowledgements{J.R.F. is supported by the U.S. National Science Foundation (NSF) Graduate Research Fellowship Program under grant No. 2139319. This research was supported in part by grant NSF PHY-2309135 to the Kavli Institute for Theoretical Physics (KITP). L.J.P is supported by a grant from the NASA Astrophysics Theory Program (ATP-80NSSC22K0725).}

\bibliography{ref}

\begin{thebibliography}{30}
\expandafter\ifx\csname natexlab\endcsname\relax\def\natexlab#1{#1}\fi

\bibitem[{Bartel(2000)}]{Bartel2000}
Bartel, N. 2000, Science, 287

\bibitem[{Bartel {et~al.}(1994)Bartel, Bietenholz, Rupen, Conway, Beasley,
  Sramek, Romney, Titus, Graham, Altunin, Jones, Rius, Venturi, Umana, Francis,
  McCall, Richer, Stevenson, Weiler, Dyk, Panagia, Cannon, Popelar, \&
  Davis}]{Bartel1994}
Bartel, N., {et~al.} 1994, Nature, 368

\bibitem[{Bartel {et~al.}(2002)Bartel, Bietenholz, Rupen, Beasley, Graham,
  Altunin, Venturi, Umana, Cannon, \& Conway}]{Bartel2002}
---. 2002, The Astrophysical Journal, 581

\bibitem[{Bietenholz {et~al.}(2005)Bietenholz, Bartel, Rupen, Beasley, Graham,
  V.I., Venturi, Umana, Cannon, \& Conway}]{Bietenholz2005}
Bietenholz, M., {et~al.} 2005, International Astronomical Union Colloquium, 192

\bibitem[{Bietenholz {et~al.}(2001)Bietenholz, Bartel, \&
  Rupen}]{Bietenholz2001}
Bietenholz, M.~F., Bartel, N., \& Rupen, M.~P. 2001, The Astrophysical Journal,
  557

\bibitem[{Bietenholz {et~al.}(2003)Bietenholz, Bartel, \&
  Rupen}]{Bietenholz2003}
---. 2003, The Astrophysical Journal, 597

\bibitem[{{Blackburn} {et~al.}(2020){Blackburn}, {Pesce}, {Johnson}, {Wielgus},
  {Chael}, {Christian}, \& {Doeleman}}]{LindyClosure}
{Blackburn}, L., {Pesce}, D.~W., {Johnson}, M.~D., {Wielgus}, M., {Chael},
  A.~A., {Christian}, P., \& {Doeleman}, S.~S. 2020, \apj, 894, 31

\bibitem[{Buss \& Fillmore(2001)}]{Buss2001}
Buss, S.~R., \& Fillmore, J.~P. 2001, ACM Transactions on Graphics, 20

\bibitem[{Chael {et~al.}(2018)Chael, Johnson, Bouman, Blackburn, Akiyama, \&
  Narayan}]{Chael2018}
Chael, A.~A., Johnson, M.~D., Bouman, K.~L., Blackburn, L.~L., Akiyama, K., \&
  Narayan, R. 2018, The Astrophysical Journal, 857, 23

\bibitem[{Chevalier \& Klein(1978)}]{Chevalier1978}
Chevalier, R.~A., \& Klein, R.~I. 1978, The Astrophysical Journal, 219

\bibitem[{EHTC {et~al.}(2019)EHTC, Akiyama, Alberdi, Alef, Asada, Azulay,
  Baczko, Ball, Balokovi{\'{c}}, Barrett, Bintley, Blackburn, Boland, Bouman,
  Bower, Bremer, Brinkerink, Brissenden, Britzen, Broderick, Broguiere,
  Bronzwaer, Byun, Carlstrom, Chael, Chan, Chatterjee, Chatterjee, Chen, Chen,
  Cho, Christian, Conway, Cordes, Crew, Cui, Davelaar, Laurentis, Deane,
  Dempsey, Desvignes, Dexter, Doeleman, Eatough, Falcke, Fish, Fomalont,
  Fraga-Encinas, Freeman, Friberg, Fromm, G{\'{o}}mez, Galison, Gammie,
  Garc{\'{i}}a, Gentaz, Georgiev, Goddi, Gold, Gu, Gurwell, Hada, Hecht,
  Hesper, Ho, Ho, Honma, Huang, Huang, Hughes, Ikeda, Inoue, Issaoun, James,
  Jannuzi, Janssen, Jeter, Jiang, Johnson, Jorstad, Jung, Karami, Karuppusamy,
  Kawashima, Keating, Kettenis, Kim, Kim, Kim, Kino, Koay, Koch, Koyama,
  Kramer, Kramer, Krichbaum, Kuo, Lauer, Lee, Li, Li, Lindqvist, Liu, Liuzzo,
  Lo, Lobanov, Loinard, Lonsdale, Lu, MacDonald, Mao, Markoff, Marrone,
  Marscher, Mart{\'{i}}-Vidal, Matsushita, Matthews, Medeiros, Menten, Mizuno,
  Mizuno, Moran, Moriyama, Moscibrodzka, M{\"{u}}ller, Nagai, Nagar, Nakamura,
  Narayan, Narayanan, Natarajan, Neri, Ni, Noutsos, Okino, Olivares,
  Ortiz-Le{\'{o}}n, Oyama, {\"{O}}zel, Palumbo, Patel, Pen, Pesce, Pi{\'{e}}tu,
  Plambeck, PopStefanija, Porth, Prather, Preciado-L{\'{o}}pez, Psaltis, Pu,
  Ramakrishnan, Rao, Rawlings, Raymond, Rezzolla, Ripperda, Roelofs, Rogers,
  Ros, Rose, Roshanineshat, Rottmann, Roy, Ruszczyk, Ryan, Rygl, S{\'{a}}nchez,
  S{\'{a}}nchez-Arguelles, Sasada, Savolainen, Schloerb, Schuster, Shao, Shen,
  Small, Sohn, SooHoo, Tazaki, Tiede, Tilanus, Titus, Toma, Torne, Trent,
  Trippe, Tsuda, van Bemmel, van Langevelde, van Rossum, Wagner, Wardle,
  Weintroub, Wex, Wharton, Wielgus, Wong, Wu, Young, Young, Younsi, Yuan, Yuan,
  Zensus, Zhao, Zhao, Zhu, Algaba, Allardi, Amestica, Anczarski, Bach,
  Baganoff, Beaudoin, Benson, Berthold, Blanchard, Blundell, Bustamente,
  Cappallo, Castillo-Dom{\'{i}}nguez, Chang, Chang, Chang, Chen, Chilson,
  Chuter, Rosado, Coulson, Crawford, Crowley, David, Derome, Dexter, Dornbusch,
  Dudevoir, Dzib, Eckart, Eckert, Erickson, Everett, Faber, Farah, Fath,
  Folkers, Forbes, Freund, G{\'{o}}mez-Ruiz, Gale, Gao, Geertsema, Graham,
  Greer, Grosslein, Gueth, Haggard, Halverson, Han, Han, Hao, Hasegawa,
  Henning, Hern{\'{a}}ndez-G{\'{o}}mez, Herrero-Illana, Heyminck, Hirota, Hoge,
  Huang, Impellizzeri, Jiang, Kamble, Keisler, Kimura, Kono, Kubo, Kuroda,
  Lacasse, Laing, Leitch, Li, Lin, Liu, Liu, Lu, Marson, Martin-Cocher,
  Massingill, Matulonis, McColl, McWhirter, Messias, Meyer-Zhao, Michalik,
  Monta{\~{n}}a, Montgomerie, Mora-Klein, Muders, Nadolski, Navarro, Neilsen,
  Nguyen, Nishioka, Norton, Nowak, Nystrom, Ogawa, Oshiro, Oyama, Parsons,
  Paine, Pe{\~{n}}alver, Phillips, Poirier, Pradel, Primiani, Raffin, Rahlin,
  Reiland, Risacher, Ruiz, S{\'{a}}ez-Mada{\'{i}}n, Sassella, Schellart, Shaw,
  Silva, Shiokawa, Smith, Snow, Souccar, Sousa, Sridharan, Srinivasan, Stahm,
  Stark, Story, Timmer, Vertatschitsch, Walther, Wei, Whitehorn, Whitney,
  Woody, Wouterloot, Wright, Yamaguchi, Yu, Zeballos, Zhang, \& Ziurys}]{EHT1}
EHTC {et~al.} 2019, ApJ, 875, L1

\bibitem[{{Filippenko} {et~al.}(1993){Filippenko}, {Matheson}, \&
  {Ho}}]{Filippenko1993}
{Filippenko}, A.~V., {Matheson}, T., \& {Ho}, L.~C. 1993, \apjl, 415, L103

\bibitem[{{Fish} {et~al.}(2016){Fish}, {Johnson}, {Doeleman}, {Broderick},
  {Psaltis}, {Lu}, {Akiyama}, {Alef}, {Algaba}, {Asada}, {Beaudoin},
  {Bertarini}, {Blackburn}, {Blundell}, {Bower}, {Brinkerink}, {Cappallo},
  {Chael}, {Chamberlin}, {Chan}, {Crew}, {Dexter}, {Dexter}, {Dzib}, {Falcke},
  {Freund}, {Friberg}, {Greer}, {Gurwell}, {Ho}, {Honma}, {Inoue}, {Johannsen},
  {Kim}, {Krichbaum}, {Lamb}, {Le{\'o}n-Tavares}, {Loeb}, {Loinard},
  {MacMahon}, {Marrone}, {Moran}, {Mo{\'s}cibrodzka}, {Ortiz-Le{\'o}n},
  {Oyama}, {{\"O}zel}, {Plambeck}, {Pradel}, {Primiani}, {Rogers}, {Rosenfeld},
  {Rottmann}, {Roy}, {Ruszczyk}, {Smythe}, {SooHoo}, {Spilker}, {Stone},
  {Strittmatter}, {Tilanus}, {Titus}, {Vertatschitsch}, {Wagner}, {Wardle},
  {Weintroub}, {Woody}, {Wright}, {Yamaguchi}, {Young}, {Young}, {Zensus}, \&
  {Ziurys}}]{Fish2016}
{Fish}, V.~L., {et~al.} 2016, \apj, 820, 90

\bibitem[{{Foreman-Mackey} {et~al.}(2013){Foreman-Mackey}, {Hogg}, {Lang}, \&
  {Goodman}}]{emcee}
{Foreman-Mackey}, D., {Hogg}, D.~W., {Lang}, D., \& {Goodman}, J. 2013, \pasp,
  125, 306

\bibitem[{{Freedman} {et~al.}(1994){Freedman}, {Hughes}, {Madore}, {Mould},
  {Lee}, {Stetson}, {Kennicutt}, {Turner}, {Ferrarese}, {Ford}, {Graham},
  {Hill}, {Hoessel}, {Huchra}, \& {Illingworth}}]{M81_Distance}
{Freedman}, W.~L., {et~al.} 1994, \apj, 427, 628

\bibitem[{{Goodman} \& {Weare}(2010)}]{MCMC}
{Goodman}, J., \& {Weare}, J. 2010, Communications in Applied Mathematics and
  Computational Science, 5, 65

\bibitem[{Heywood {et~al.}(2009)Heywood, Blundell, Kl{\"{o}}ckner, \&
  Beasley}]{Heywood2009}
Heywood, I., Blundell, K.~M., Kl{\"{o}}ckner, H.~R., \& Beasley, A.~J. 2009,
  Monthly Notices of the Royal Astronomical Society, 392

\bibitem[{{Johnson} {et~al.}(2015){Johnson}, {Fish}, {Doeleman}, {Marrone},
  {Plambeck}, {Wardle}, {Akiyama}, {Asada}, {Beaudoin}, {Blackburn},
  {Blundell}, {Bower}, {Brinkerink}, {Broderick}, {Cappallo}, {Chael}, {Crew},
  {Dexter}, {Dexter}, {Freund}, {Friberg}, {Gold}, {Gurwell}, {Ho}, {Honma},
  {Inoue}, {Kosowsky}, {Krichbaum}, {Lamb}, {Loeb}, {Lu}, {MacMahon},
  {McKinney}, {Moran}, {Narayan}, {Primiani}, {Psaltis}, {Rogers}, {Rosenfeld},
  {SooHoo}, {Tilanus}, {Titus}, {Vertatschitsch}, {Weintroub}, {Wright},
  {Young}, {Zensus}, \& {Ziurys}}]{Johnson2015}
{Johnson}, M.~D., {et~al.} 2015, Science, 350, 1242

\bibitem[{Johnson {et~al.}(2019)Johnson, Lupsasca, Strominger, Wong, Hadar,
  Kapec, Narayan, Chael, Gammie, Galison, Palumbo, Doeleman, Blackburn,
  Wielgus, Pesce, Farah, \& Moran}]{Johnson2019}
Johnson, M.~D., {et~al.} 2019, arXiv e-prints, arXiv:1907.04329

\bibitem[{Khumpasee {et~al.}(2024)Khumpasee, Niwitpong, \& Niwitpong}]{HPDI}
Khumpasee, W., Niwitpong, S.-A., \& Niwitpong, S. 2024, Symmetry, 16, 1488

\bibitem[{{Kundu} {et~al.}(2019){Kundu}, {Lundqvist}, {Sorokina},
  {P{\'e}rez-Torres}, {Blinnikov}, {O'Connor}, {Ergon}, {Chandra}, \&
  {Das}}]{Kundu2019}
{Kundu}, E., {et~al.} 2019, \apj, 875, 17

\bibitem[{{Marcaide} {et~al.}(1995){Marcaide}, {Alberdi}, {Ros}, {Diamond},
  {Schmidt}, {Shapiro}, {Baath}, {Davis}, {de Bruyn}, {El{\'o}segui},
  {Guirado}, {Jones}, {Krichbaum}, {Mantovani}, {Preston}, {Ratner}, {Rius},
  {Rogers}, {Schilizzi}, {Trigilio}, {Whitney}, {Witzel}, \&
  {Zensus}}]{Marcaide1995SHELL}
{Marcaide}, J.~M., {et~al.} 1995, \nat, 373, 44

\bibitem[{Marcaide {et~al.}(1995)Marcaide, Alberdi, Ros, Diamond, Shapiro,
  Guirado, Jones, Krichbaum, Mantovani, Preston, Rius, Schilizzi, Trigilio,
  Whitney, \& Witzel}]{Marcaide1995}
Marcaide, J.~M., {et~al.} 1995, Science, 270

\bibitem[{Marcaide {et~al.}(1997)Marcaide, Alberdi, Ros, Diamond, Shapiro,
  Guirado, Jones, Mantovani, P{\'{e}}rez-Torres, Preston, Schilizzi, Sramek,
  Trigilio, {Van Dyk}, Weiler, \& Whitney}]{Marcaide1997}
---. 1997, The Astrophysical Journal, 486

\bibitem[{Marcaide {et~al.}(2009)Marcaide, Mart{\'{i}}-Vidal, Alberdi,
  P{\'{e}}rez-Torres, Ros, Diamond, Guirado, Lara, Shapiro, Stockdale, Weiler,
  Mantovani, Preston, Schilizzi, Sramek, Trigilio, {Van Dyk}, \&
  Whitney}]{Marcaide2009}
---. 2009, Astronomy and Astrophysics, 505

\bibitem[{{Mart{\'\i}-Vidal} {et~al.}(2024){Mart{\'\i}-Vidal}, {Bj{\"o}rnsson},
  {P{\'e}rez-Torres}, {Lundqvist}, \& {Marcaide}}]{Vidal2024}
{Mart{\'\i}-Vidal}, I., {Bj{\"o}rnsson}, C.~I., {P{\'e}rez-Torres}, M.~A.,
  {Lundqvist}, P., \& {Marcaide}, J.~M. 2024, \aap, 691, A171

\bibitem[{{Ripero} {et~al.}(1993){Ripero}, {Garcia}, {Rodriguez}, {Pujol},
  {Filippenko}, {Treffers}, {Paik}, {Davis}, {Schlegel}, {Hartwick}, {Balam},
  {Zurek}, {Robb}, {Garnavich}, \& {Hong}}]{Disc_93J}
{Ripero}, J., {et~al.} 1993, \iaucirc, 5731, 1

\bibitem[{Thompson {et~al.}(2001)Thompson, Moran, \& Swenson}]{Thompson2001}
Thompson, A.~R., Moran, J.~M., \& Swenson, G.~W. 2001, {Interferometry and
  Synthesis in Radio Astronomy}

\bibitem[{Vehtari {et~al.}(2021)Vehtari, Gelman, Simpson, Carpenter, \&
  B{\"u}rkner}]{ImprovedRhat}
Vehtari, A., Gelman, A., Simpson, D., Carpenter, B., \& B{\"u}rkner, P.-C.
  2021, Bayesian Analysis, 16, 667

\bibitem[{Weiler {et~al.}(2007)Weiler, Williams, Panagia, Stockdale, Kelley,
  Sramek, {Van Dyk}, \& Marcaide}]{Weiler2007}
Weiler, K.~W., Williams, C.~L., Panagia, N., Stockdale, C.~J., Kelley, M.~T.,
  Sramek, R.~A., {Van Dyk}, S.~D., \& Marcaide, J.~M. 2007, The Astrophysical
  Journal, 671

\end{thebibliography}


\end{document}